\documentclass{aa}
\usepackage{graphicx}
\usepackage{txfonts}
\usepackage{color}
\begin{document}
   \title{Diffusive convective overshoot in core He-burning
          intermediate mass stars. I: the LMC metallicity}
   
   \author{P. Ventura\inst{1}, M. Castellani\inst{1}
          \and
          C.W. Straka\inst{2} 
          }

   \offprints{P. Ventura}

   \institute{Osservatorio Astronomico di Roma
              Via Frascati 33 00040 MontePorzio Catone - Italy\\
              \email{ventura@mporzio.astro.it, 
                     m.castellani@mporzio.astro.it}             
              \and
              Department of Astronomy, Yale University,
              P.O. Box 208101, New Haven, CT 06520-8101 \\
              \email{straka@astro.yale.edu}                
                               }

   \date{Received ... ; accepted ...}

   \abstract{We present detailed evolutionary calculations focused
             on the evolution of intermediate mass stars
             with $3M_{\odot} \leq M \leq 9M_{\odot}$ of metallicity
             typical of the Large Magellanic Cloud (LMC), 
             i.e. $Z=0.008$. We compare carefully
             the models calculated by adopting a diffusive 
             scheme for chemical mixing, in which nuclear burning and 
             mixing are self-consistently coupled, 
             while the eddy velocities beyond the formal convective core
	     boundary are treated to decay exponentially, and those
             calculated with the traditional 
             instantaneous mixing approximation. 
             We find that: i) the physical and chemical behaviour of the
             models during the H-burning phase is independent of the scheme 
             used for the treatment of mixing inside the CNO burning core;
             ii) the duration of the He-burning phase relative 
             to the MS phase is systematically longer in the diffusive models,
             due to a slower redistribution of helium to the core from the 
             outer layers; iii) the fraction of time spent in the blue part 
             of the clump, compared to the stay in the red, is larger in the 
             diffusive models. The differences described in points ii) and iii)
             tend to vanish for 
             $M > 6M_{\odot}$. In terms of the theoretical interpretation of 
             an open cluster stellar population, the differences introduced 
             by the use of a self-consistent scheme for mixing in the core
	     with adjacent exponential decay
             are relevant for ages in the range 80 Myr $< t <$ 200 Myr.
             These results are robust, since they are
	     insensitive to the choice of the free-parameters regulating
	     the extension of the extra-mixing region.

   }

   \maketitle
%

\section{Introduction}
The stars of intermediate mass (hereinafter IMS) are by 
definition those which are massive enough to undergo 
helium burning, but which never reach conditions
in the central regions leading to $^{12}C$ ignition: soon
after the end of helium burning in the core they undergo
a phase of thermal pulses, during which the star for
most of the time is supported energetically by a CNO
burning shell, with periodic and thermally unstable
ignition of $3\alpha$ reactions in a stellar layer
above the CO core (Schwarzschild \& Harm 1965, 1967;
Iben 1975, 1976); eventually, mass loss determines 
a general cooling of the structure and a later 
evolution as a CO white dwarf.

During the last decades the interest towards the evolution
of this class of objects has raised, because they were
suggested responsible for chemical 
anomalies observed at the surface of globular cluster
stars (Gratton et al. 2004): the idea behind 
this hypothesis is that an 
early generation of IMS evolved on short time-scales
within the cluster, and polluted the interstellar
medium with nuclearly processed material, so that a
new generation of stars formed in an environment
which, on a chemical point of view, might show the
anticorrelations currently observed (Cottrell 
\& Da Costa 1981; D'Antona et al. 1983; Ventura et 
al. 2001, 2002).

During the two phases of nuclear burning in the
central regions, the IMS develop extended convective 
cores, the sizes of which increase with mass. On
the theoretical side the duration of these phases,
and the path followed by the tracks in the HR
diagram, are uncertain because of the 
unknown behaviour of the convective 
eddies as they approach the 
formal border of the core, the location of which is 
found via the traditional Schwarzschild criterion.
Unfortunately, this latter criterion allows to locate the point
where buoyancy vanishes, but the distance 
(due to inertia) convective eddies travel beyond 
the border cannot be inferred.

Traditionally, in the evolution codes, this 
extra-mixing region, which is formally radiative,
has been simulated by assuming that the zone
out of the formally convective core which is
fully homogenised has a geometrical width which
is parametrised as $\alpha H_p^b$, where $H_p^b$
is the pressure scale height at the border of
the core, and $\alpha$ is a free parameter which
is calibrated to reproduce the MS  of open
clusters (instantaneous mixing). 
Detailed comparisons of theoretical 
predictions with photometric studies of open
clusters have shown that a value of $\alpha \approx 0.2$
is needed for IMS, at least for $M > 2M_{\odot}$
(see e.g. Maeder \& Meynet 1991; 
Stothers \& Chin 1992).

Alternatively, the extra-mixing beyond the formal
convective border may be simulated by assuming 
that the velocities of the convective eddies
decay exponentially, with a more physical description,
in which mixing of chemicals and nuclear burning 
are self-consistently coupled (diffusive scheme)  
(Deng at al. 1996a,b; Herwig et al. 1997;
Ventura et al. 1998).

In a recent paper focused on reproducing the stellar 
population of the clump region of the LMC cluster
NGC 1866, Ventura \& Castellani (2005, hereinafter 
VC05) showed that using a diffusive approach 
within the convective core of the IMS leads to 
a longer stay of the evolutionary
tracks in the blue part of the HR diagram during the
He-burning phases; this, in turn, helped in reconciling
the observed scenario for NGC 1866 (in which the ratio
$B/R$ between stars burning helium in the blue and red
region of the clump is $\approx 1$) (Testa et al. 1999)
with the theoretical 
predictions (a $B/R \approx 0.8$ is found with the
diffusive case, to be compared to $B/R \approx 0.4$
if an instantaneous scheme is adopted). 
The particular scheme adopted to deal with chemical
mixing inside the He-burning convective core turned
out to be by far the most relevant physical input
influencing the distribution of stars in the clump
region of the HR diagram, when compared to
other model uncertainties of, e.g., the assumptions made 
upon the extension of the extra-mixing region, the cross 
sections used for the $^{12}C+\alpha$ reaction, and the 
convective model adopted to find out the temperature gradient 
within regions unstable to convection.
A further important result of that study was that
the whole He-burning phase was longer in the diffusive
models, so that larger $t_{He}/t_{H}$ ratios are
expected.

The conclusions of the afore mentioned paper might
hold in principle just for the range of masses relevant
for the study of the clump region of NGC 1866
(i.e. $\approx 4 - 4.5M_{\odot}$); the idea behind this
paper is therefore to extend the above analysis to 
a wider range of masses, ranging from the minimum mass
undergoing an overall contraction during the core
He-burning phase leading to a meaningful excursion of
the track to the blue (i.e. $3M_{\odot}$) through the
whole range of IMS stars. We actually decided to extend 
our analysis a bit further, up to $9M_{\odot}$.
We postpone to a forthcoming paper a wider analysis,
including the effects of metallicity.

We compare carefully the results obtained with the
instantaneous and the diffusive scheme for chemical
mixing for all the masses involved. We also test
for the correct distribution of stars which should 
be expected in the clump region of the HR diagram
at different ages (hence, mass in the clump).
Basically, our aim is to understand in which
cases (hence, for which masses and ages) the use
of a diffusive scheme to deal with nuclear
burning and chemical mixing within the He-burning
cores of IMS is mandatory for a reliable estimate
of the relative duration of the various evolutionary
phases.

We provide a description of the main physical inputs
of the code used to calculate the evolutionary sequences
in Sect.2. In Sect.3 we present and discuss the main
properties and the various evolutionary phases of the
diffusive models, focusing in particular on the
He-burning phase; we evaluate the reliability of our
results for what concerns the description of the 
kinematic behaviour of the 
convective eddies in the proximity of the border 
of the core, comparing the velocities found via the use
of the local FST convective scheme with those obtained with
a more physically-sound non-local treatment. Finally,
we discuss the differences between the diffusive models and 
those calculated with the instantaneous mixing approximation 
in Sect.4, where we also suggest the range of masses and
ages for which the use of the diffusive approach is 
recommended.  

\section{The stellar evolution code}
All the evolutions presented and discussed in this
paper were calculated by means of the code ATON2.1,
a full description of which can be found in
Ventura et al. (1998), with the last updates
given in Ventura \& D'Antona (2005); the interested
reader may find in the two above papers all the
details concerning the numerical structure of the
code, and the micro- and macro-physics input.

Here we want just to recall some of the physical
inputs most relevant for this work:

\begin{itemize}
\item{The temperature gradient within instability
regions can be found either by the traditional
Mixing Length Theory (MLT, B\"ohm-Vitense 1958;
Vitense 1953), or via the Full Spectrum
of Turbulence (FST) model (Canuto \& Mazzitelli
1991).}

\item{Within regions unstable to convective motions,
it is possible to deal with nuclear burning either
by assuming that mixing is instantaneous (so that
the whole convective region is assumed to be always 
chemically homogenised), or by
treating simultaneously nuclear burning and mixing
of chemicals; in this latter case we solve for 
each of the nuclear species included in the network
the diffusion equation (Cloutman \& Eoll 1976):

\begin{equation}
$$
  \left( {dX_i\over dt} \right)=\left( {\partial X_i\over \partial t}\right)_{\rm nucl}+
  {\partial \over \partial m_r}\left[ (4\pi r^2 \rho)^2 D {\partial X_i
  \over \partial m_r}\right]  \label{diffeq}
$$
\end{equation}

stating mass conservation of element $i$. The diffusion 
coefficient $D$ is taken as

\begin{equation}
$$
D={1\over 3}vl
$$
\end{equation}
where $v$ is the convective velocity and $l$ is the convective scale 
length.

Within this diffusive framework it is necessary to 
specify the way with which convective velocities decay 
outside the convective boundaries (Deng at al. 1996a,b; 
Herwig et al. 1997; Ventura et al. 1998). 
In agreement with Xiong (1985) and Grossman (1996)  
and supported by the numerical simulations by Freytag 
et al. (1996), we assume that convective velocities 
decay exponentially outside the formal convective 
boundary as:

\begin{equation}
$$
v=v_b exp \pm \left( {1\over \zeta f_{\rm thick}}ln{P\over P_b}\right) 
$$
\end{equation}

\noindent
where $v_b$ and $P_b$ are, respectively, turbulent 
velocity and pressure at the convective boundary, P 
is the local pressure, $\zeta$ a free parameter 
connected with the e-folding distance of the decay, 
and $f_{\rm thick}$ is the thickness of the convective 
regions in fractions of $H_p$.  

In the instantaneous case the geometrical width of 
the extra-mixing region is parametrised as 
$l_{ov}=\alpha H_p$. The physical meanings of the
parameters $\zeta$ (diffusive models) and $\alpha$ 
(instantaneous) are strongly different (see the
detailed discussion above this argument in VC05), but,
in terms of the duration and extension of the MS
phase, we find that $\alpha=0.2$ is equivalent to
$\zeta=0.03$. These are the values of $\alpha$ and
$\zeta$ which will be used for our analysis.  
}

\item{All the nuclear cross-sections are taken from
Angulo et al. (1999), with the only exception of the
$^{12}C(\alpha,\gamma)^{16}O$ reaction, for which
we employ the more accurate determination by Kunz et 
al. (2002).}

\end{itemize}

%
   \begin{table}
      \caption[]{Physical properties of the diffusive models.}
         \label{physics}
       $$ 
           \begin{array}{c c c c c c c c c}
            \hline
            \noalign{\smallskip}
            M_{ZAMS} & t(H)^a & M_{core,H}^b & M_{1st d.up}^d & \log({L_{tip}\over L_{\odot}}) & M_{core,He}^c & t(He)^a & t_{blue}^a & t_{red}^a \\
            \noalign{\smallskip}
            \hline
            \noalign{\smallskip}
            3.0 & 363 & 0.63 & 0.53 & 2.75 & 0.20 & 75  & 0   & 75     \\
            4.0 & 179 & 0.97 & 0.84 & 3.11 & 0.30 & 25  & 16  & 9     \\
            5.0 & 106 & 1.29 & 1.18 & 3.43 & 0.42 & 13  & 8   & 4     \\
            6.0 & 72  & 1.64 & 1.52 & 3.70 & 0.56 & 7.3 & 5.3 & 2     \\
            7.0 & 52  & 2.00 & 1.89 & 3.93 & 0.72 & 4.9 & 3.6 & 1.3     \\
            8.0 & 40  & 2.38 & 2.29 & 4.12 & 0.92 & 3.6 & 2.6 & 0.9     \\
            9.0 & 32  & 2.81 & 2.71 & 4.29 & 1.17 & 2.9 & 2.2 & 0.6     \\

            \noalign{\smallskip}
            \hline
         \end{array}
     $$ 
\begin{list}{}{}
\item[$^{\mathrm{a}}$] Times are expressed in Myr.
\item[$^{\mathrm{b}}$] Maximum extension (in solar masses) 
of the convective core during H-burning.
\item[$^{\mathrm{c}}$] Maximum extension (in solar masses) 
of the convective core during He-burning.
\item[$^{\mathrm{d}}$] Mass coordinate (in solar masses) 
of the innermost layer reached during the first dredge-up.
\end{list}
   \end{table}

\section{The evolution of the diffusive models}
The diffusive models were
calculated by adopting a chemical composition typical
of the LMC, i.e. $Z=0.008$ and $Y=0.25$ (Brocato et 
al. 2003; Barmina et al. 2002). For all
the other elements we assumed solar-scaled abundances.

Convection was treated according to the FST framework:
we recall, however, that one of the main results of
VC05 was that the convective model does influence only
the colour and the slope of the giant branch in the
HR diagram, but neither the thermodynamic structure of
the convective cores (both during hydrogen and helium
burning), nor the duration of the different evolutionary
phases.

\begin{figure}
\includegraphics[width=8cm]{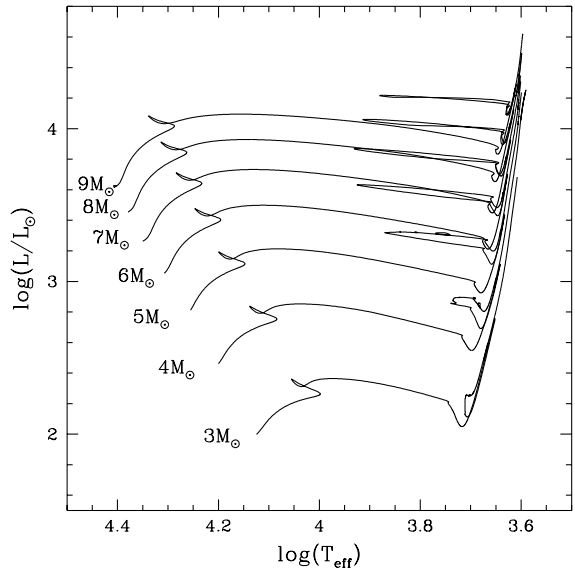}
\caption{Stellar tracks on the HR diagram related
         to the evolution of the intermediate mass 
         diffusive models. Labels indicate
         the stellar mass. The pre-MS evolution was
         omitted for clarity reasons.}
         \label{HRdiagram}%
\end{figure}

In agreement with VC05, we used the
diffusive scheme to
couple nuclear burning and chemical mixing; the 
coefficient $\zeta$ for the exponential decay of 
velocities out of the convective cores was set
to $\zeta=0.03$. No extra-mixing from the base of
the convective envelope was considered.

Breathing pulses were suppressed,
by preventing any growth of the convective core
once the central abundance of helium drops below
0.05.

Fig.~\ref{HRdiagram} shows the tracks followed by 
the models considered on the theoretical plane
$\log(T_{eff}) - \log(L/L_{\odot}$). The evolutions
were calculated starting from the early evolutionary
stages, but the pre-MS phase was omitted for 
clarity reasons.

The main properties of the IMS models are reported
in Tab.~\ref{physics}. 

\begin{figure*}
\centering{
\includegraphics[width=8cm]{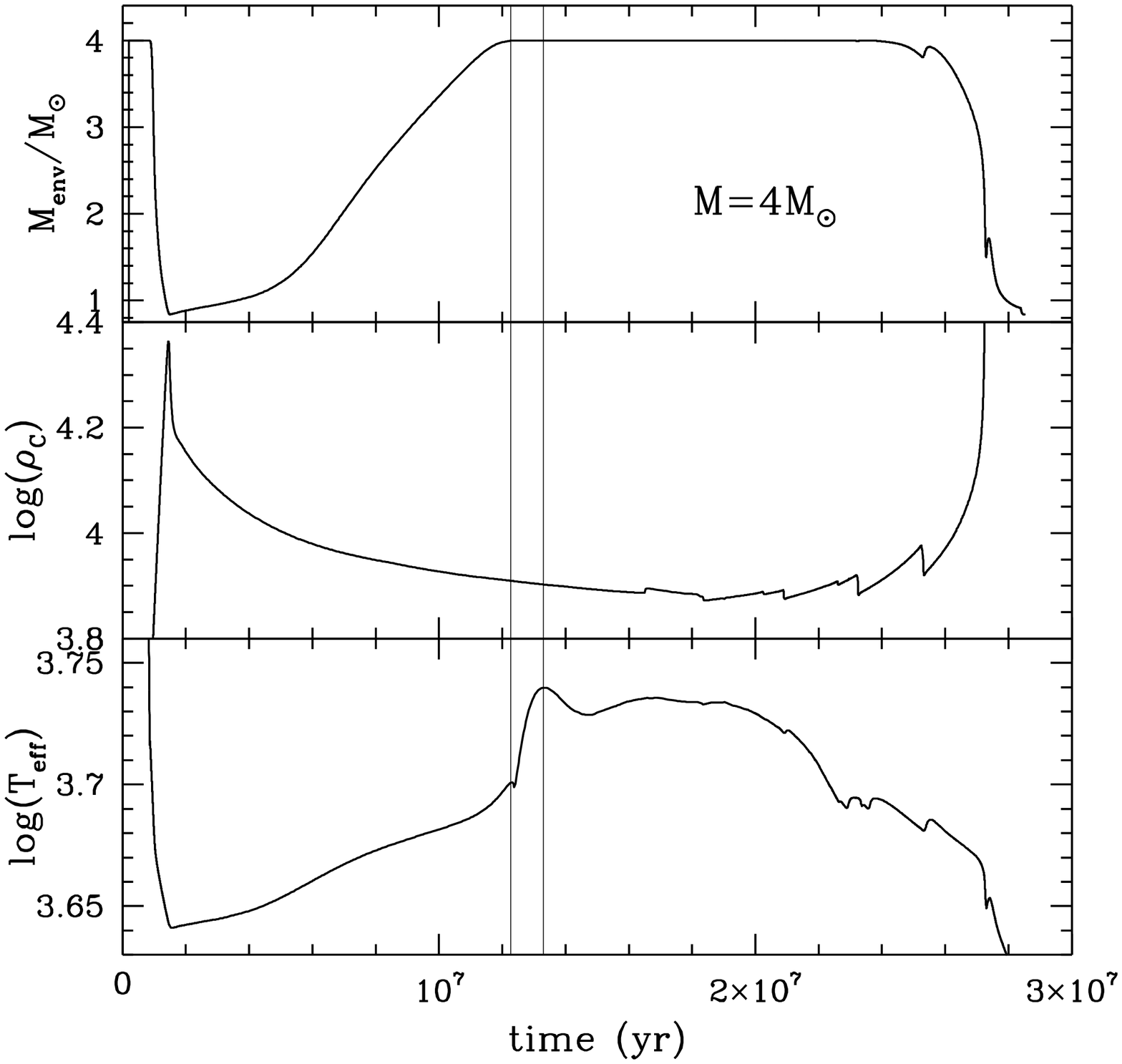}
\includegraphics[width=8cm]{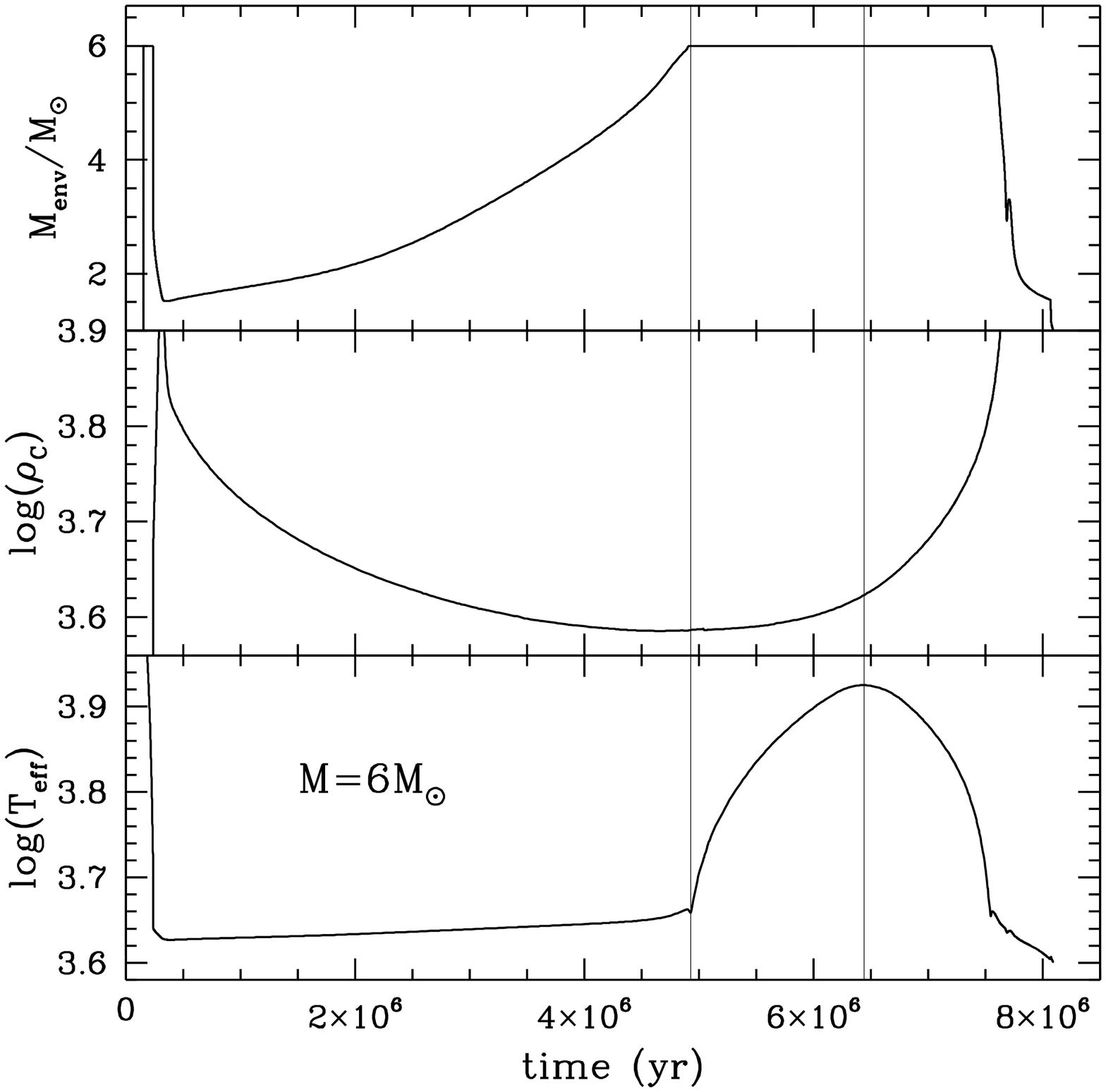}
           }
\caption{The temporal evolution of some physical
         quantities related to the evolution of two
         stellar models of mass $4M_{\odot}$ (left)
         and $6M_{\odot}$ (right). For each mass
         the variation with time of
         the base of the convective envelope (top),
         central density (middle) and effective
         temperature (bottom) is reported. The thin 
         vertical lines in both panels indicate the 
         time when the surface 
         convection is extinguished (which is coincident
         with the beginning of the excursion
         of the track to the blue) and when the bluest
         point on the HR diagram is reached.}
   \label{menv}
\end{figure*}

During the MS phase all models attain temperatures
in the central regions sufficiently high to convert
protons to helium via the CNO cycle; this in turn 
favours the formation of a convective core. As 
can be seen in the 3rd column of 
Tab.~\ref{physics}, the maximum extension 
(in mass) of the core is an increasing
function of the total mass, ranging from 
$0.63M_{\odot}$ for the $3M_{\odot}$ model to 
$2.81M_{\odot}$ for $M=9M_{\odot}$. 
The duration of the whole phase of hydrogen 
burning ($t_H$) decreases with mass: we find 
$t_H=363$ Myr for $M=3M_{\odot}$, down to 
$t_H=32$ Myr for $M=9M_{\odot}$.

As can be seen from Fig.~\ref{HRdiagram} the stars,
soon after central hydrogen exhaustion, undergo a phase
of general expansion of the outer layers, while CNO
burning continues within a shell laying above the 
contracting core (Iben 1993). The decrease of the 
effective temperature favours the inwards 
penetration of the convective envelope,
which reaches layers previously touched 
(at least partially) by CNO burning, in what is 
commonly known as the first dredge-up episode. 
Column 6 of Tab.~\ref{physics}
reports the mass coordinate of the innermost layer
reached by the external convection during this process:
we see that the mass of the whole external envelope
at the maximum penetration of the surface convection is
increasing with mass, reaching a maximum value of
$\delta M \approx 8M_{\odot}$ for the $9M_{\odot}$ model.

As the CNO shell moves outwards the luminosity 
increases, until the temperature in the centre
reaches values sufficiently high to favour helium burning 
via the $3\alpha$ reactions. We see that the 
temperature at which helium burning begins is slightly
dependent on mass, ranging from $T_c \approx 120 \times 10^6$
K for $M=3M_{\odot}$ to $T_c \approx 148 \times 10^6$ K
for $M=9M_{\odot}$. The beginning of core He-burning is
accompanied by an expansion of the central regions,
with the subsequent cooling of the CNO shell, so that
the stellar luminosity reaches a tip value (reported for each 
model in column 5 of Tab.~\ref{physics}) and then decreases.

All the main steps related to the evolution of
IMS through the helium burning phase, and the 
occurrence of the blue loops on the HR diagram, 
are well documented in the literature (see e.g
Chiosi et al. 1992), and will
not be repeated here. We only want to recall
the main factors triggering the excursion of
the tracks towards the blue and the relative 
duration of the stay in this region (VC05):

\begin{itemize}
\item{The overall contraction leading to the
bluewards excursion starts as soon as the surface
convection disappears.}

\item{The tracks stay in the blue until the
scarcity of helium in the central regions forces
the core to contract, with the consequent
expansion of the external layers.}

\item{The duration of the blue phase is therefore
connected with the helium still left
in the core when the blue region of the clump
is reached: the lower the helium mass fraction,
the shorter is the stay in the blue.}

\end{itemize}
\noindent
A rapid glance at columns 8 and 9 of 
Tab.~\ref{physics} shows that the percentage
of time spent in the blue compared to
the one spent in the red is decreasing
with mass, ranging from $t_{blue}/t_{red}
\approx 0.7$ for the $4M_{\odot}$ model to
$t_{blue}/t_{red} \approx 0.25$ for 
$M=9M_{\odot}$. We also note from 
Fig.~\ref{HRdiagram} that the extension of
the loops decreases for $M>7M_{\odot}$.

The decrease with mass of the fraction of the total
He-burning time spent in the blue can be understood by 
looking at Fig.~\ref{menv}, where we report the temporal
evolution of the He-burning phase of two models with 
masses $M=4M_{\odot}$ (left) and $M=6M_{\odot}$ (right)
in terms of the variations with time of the base of the 
convective envelope ($M_{env}$), central density ($\rho_C$), 
and effective temperature ($T_{eff}$).

\begin{figure}
\includegraphics[width=8cm]{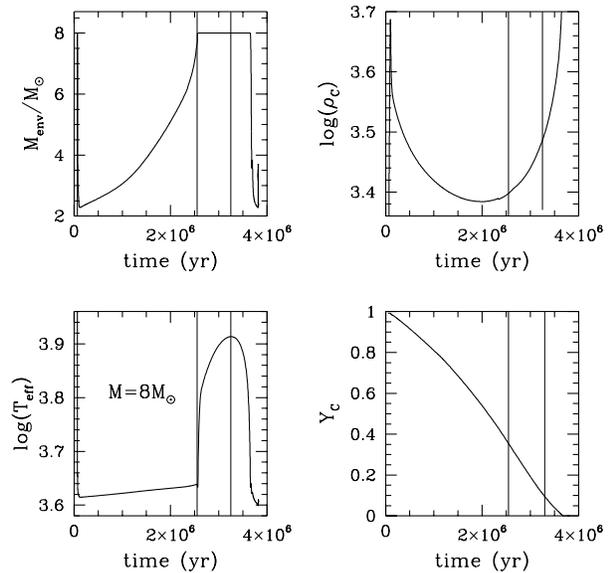}
\caption{The variation with time of some physical
         and chemical quantities of the evolution of
         a $8M_{\odot}$ diffusive model: we show the
         mass of the base of the surface convection
         (top-left panel), the central density 
         (top-right), the effective temperature
         (bottom-left) and the central mass fraction
         of helium (bottom-right). The thin vertical
         lines have the same meaning as in
         Fig.~\ref{menv}.}
         \label{8msun}%
\end{figure}

We see that the excursion to the blue, which can be identified by a 
rapid change of slope of the variation with time of $T_{eff}$, starts
as soon as the surface convection is extinguished. As can be seen by 
comparing the left and right top panels, the mass of the envelope at 
the beginning of the He-burning phase is larger for the more 
massive models, so that by the time that the surface convection vanishes 
the helium left in the centre is lower: the consequence is that 
for larger masses, as soon as the track reaches the bluest point, the 
contraction of the core, which can be seen by the increase of the central 
density shown in the middle panels of Fig.~\ref{menv},  follows shortly
afterwards, hence the track returns quickly to the red. If we compare 
the evolution of the central densities of the two models, we see that 
in the $4M_{\odot}$ model the central regions are still expanding when 
the maximum $T_{eff}$ is reached, while in the $6M_{\odot}$ model the 
core has just started contracting.

Even the shorter extension of the loops of the largest masses which 
can be seen in Fig.~\ref{HRdiagram} can be explained on the basis of 
the percentage of helium in the core when the surface convection
disappears. In Fig.~\ref{8msun} we show the details of the evolution 
of the $8M_{\odot}$ model, during the core He-burning phase. In this
case we also show the variation of the central helium (right-bottom panel). 
The vertical thin lines mark, respectively, the stage when the overall 
contraction begins, and when the bluest point of the track in the HR 
diagram is reached. We see that when the surface envelope disappears 
the helium left in the core is less than $40\%$, and more important,  
when the bluest point is reached $Y_C \approx 0.1$, and the core contraction 
has already begun (see the right-top panel). In this case, therefore, the
scarcity of helium triggers the core contraction while the track was 
still moving to the blue, with a consequent shorter extension of the loop. 
 
\begin{figure*}
\centering{
\includegraphics[width=8cm]{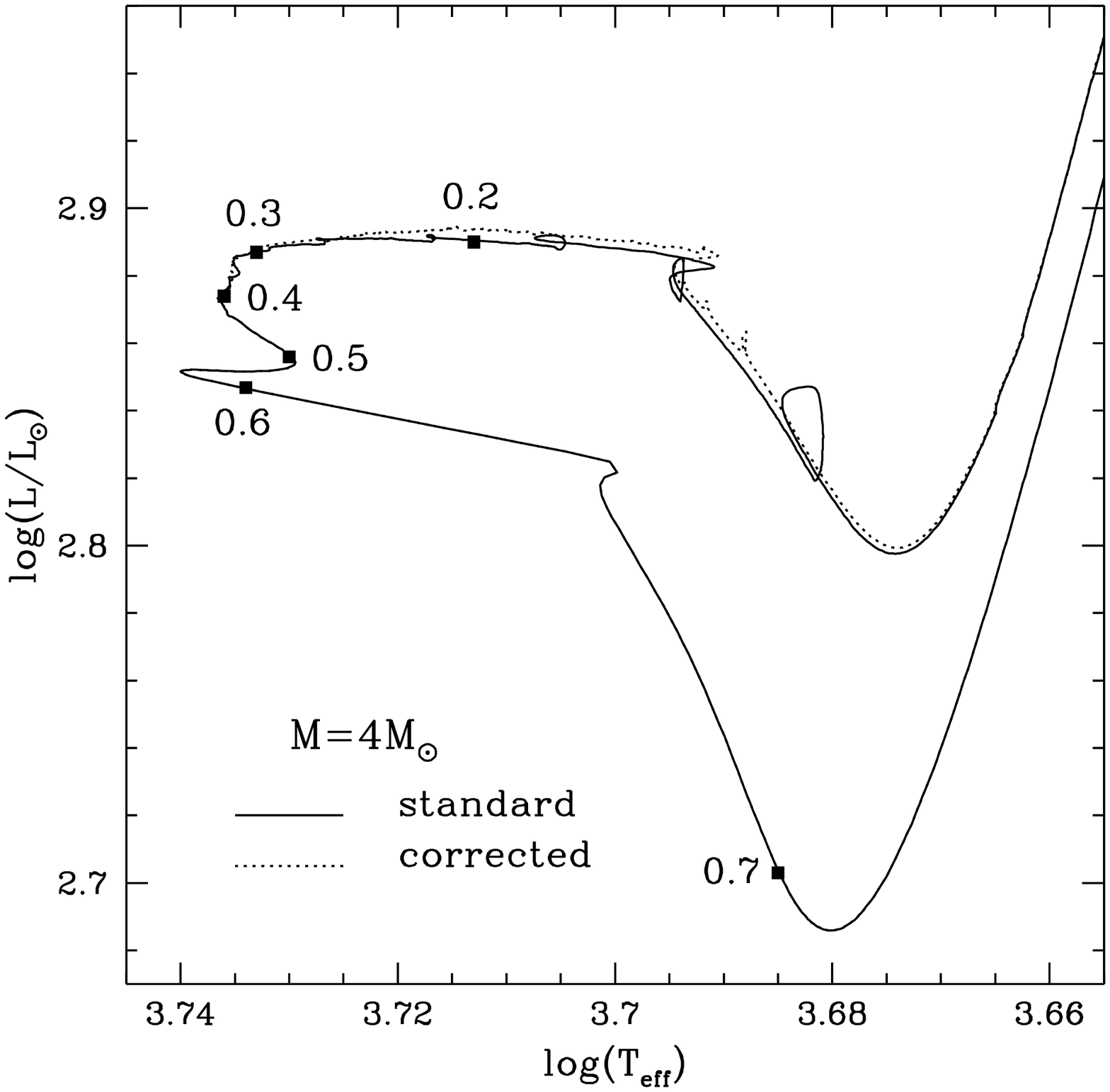}
\includegraphics[width=8cm]{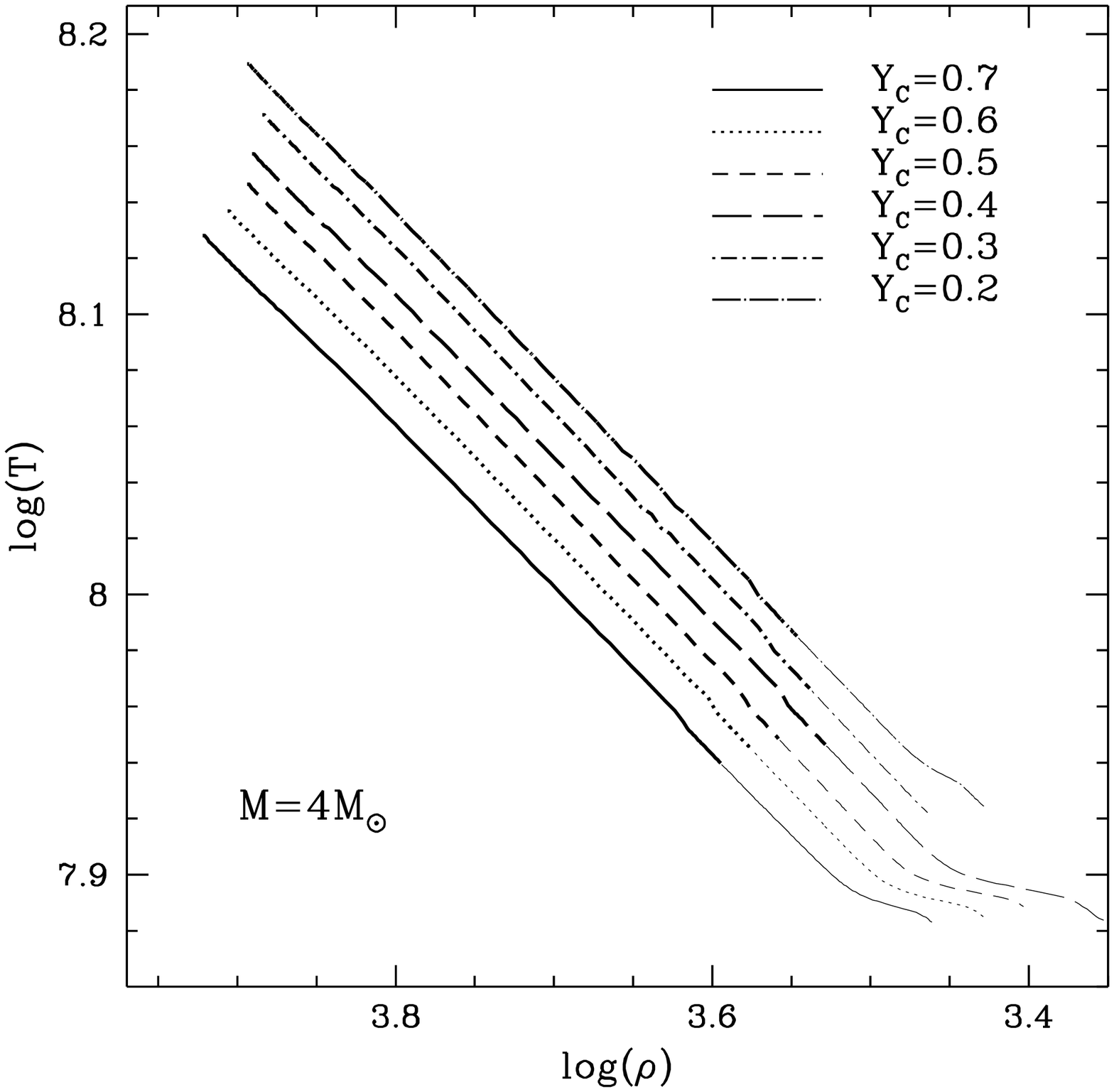}
           }
\caption{{\bf Left}: The path followed in the HR diagram of 
         a $4M_{\odot}$ diffusive model during the 
         He-burning phase (solid line); the labels on the
         track indicate the corresponding values of the
         central helium mass fraction. The dotted line
         corresponds to a model where any convection
         zone in the proximity of the core was ignored
         in the calculation of the velocity profile
         following the method described in the text.
         {\bf Right}: The internal structure of the 
         He-burning core of the same $4M_{\odot}$ model
         reported on the left panel in the $\log(\rho) -
         \log(T)$ plane. The structures refer to the
         evolutionary stages marked with square dots 
         on the track in
         the left panel. The thick parts refer to the
         zone which is formally convective according to
         the Schwarzschild criterion. 
         }   
   \label{4morot}
\end{figure*}

\subsection{Semiconvection}
Fig.~\ref{HRdiagram} (and in more details
also the left panel of Fig.~\ref{4morot})
shows that the tracks of the lowest masses of 
our sample are characterised by some loops of 
minor extension on the HR diagram; this happens 
only after the bluest point is reached,
when the helium abundance dropped to values
$Y_C \leq 0.2$. The high mass tracks don't show
such irregularities.

As expected, we could verify that the occurrence
of such loops, with a temporary increase of both
luminosity and effective temperature, is 
connected with a sudden increase of the helium mass 
fraction in the whole core, with the consequent rapid 
increase of the helium burning rate in the central 
regions. Traces of these episodes can also be seen 
in the temporal
variation of the central density of the $4M_{\odot}$
model reported in Fig.~\ref{menv} (left-middle
panel). These irregularities are due to the formation 
of a small convective region in the proximity of 
the border of the convective core, triggered by 
the appearance of a minimum in the profile of the 
radiative logarithmic temperature gradient $\nabla_{rad}$
(Castellani et al. 1971a,b; Castellani et al. 1985;
Alongi et al. 1993). We verified that any
change in the temporal or spatial resolution used
in calculating the evolutions may only shift sligthly
the location of these loops in the HR diagram,
but cannot prevent these irregularities, because
the small convective shell is formed in any case.

We went into the details of the evolution of the
$4M_{\odot}$ model, which is the typical mass
showing such irregular behaviour. 
Fig.~\ref{4morot} shows the path followed by the
model in the HR diagram focused on the clump
region (solid track, left panel), and the 
evolution of the thermodynamic structure of the 
core in the $\log(\rho) - \log(T)$ plane at 
some selected evolutionary stages (right panel), 
corresponding to some fixed values of the central 
helium abundance, and marked by full dots in
the left panel. We note that while the core
temperatures increase for the whole helium
burning phase, the densities first decrease 
and then begin to increase when $Y_C < 0.3$.

\begin{figure*}
\centering{
\includegraphics[width=8cm]{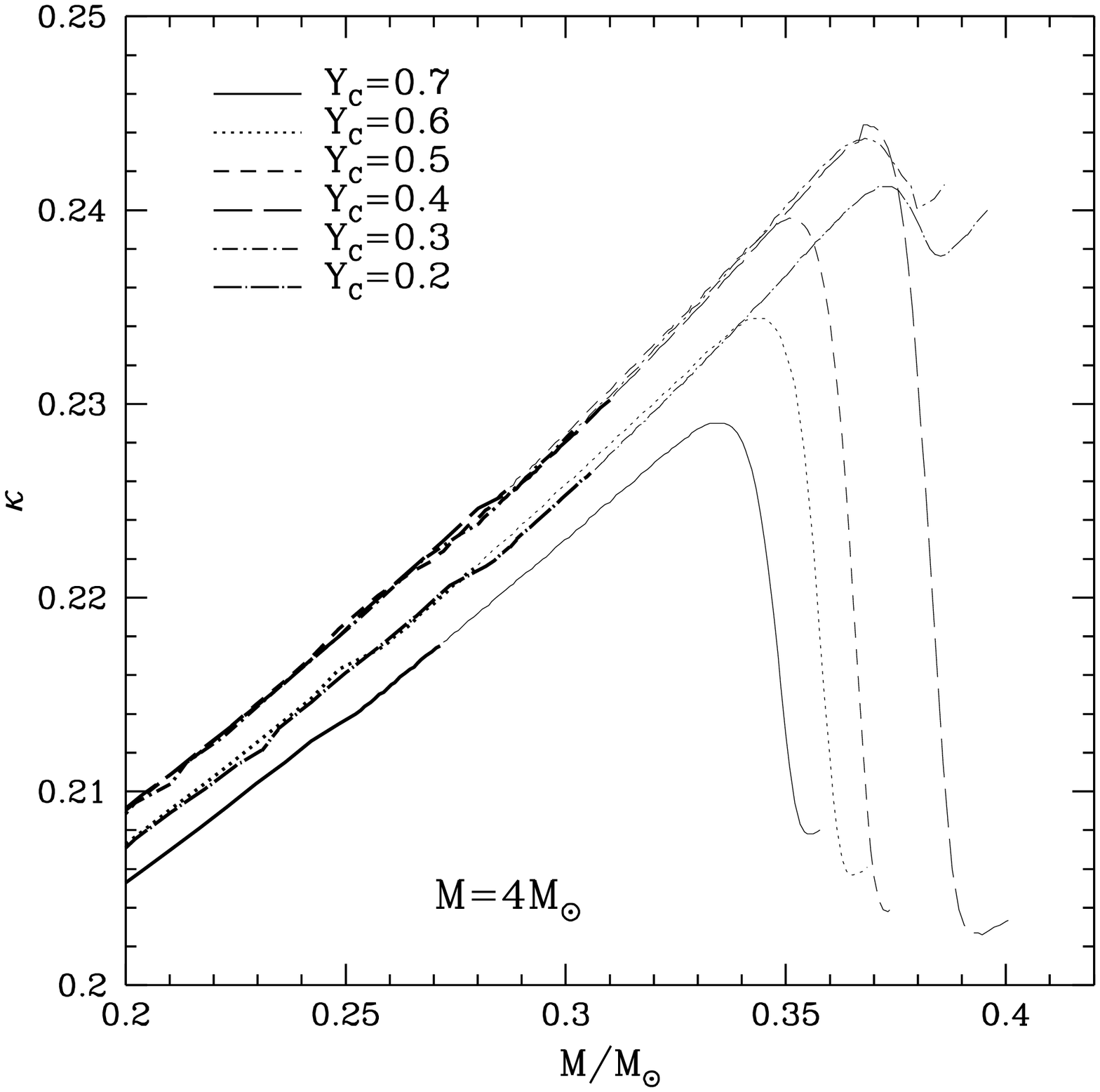}
\includegraphics[width=8cm]{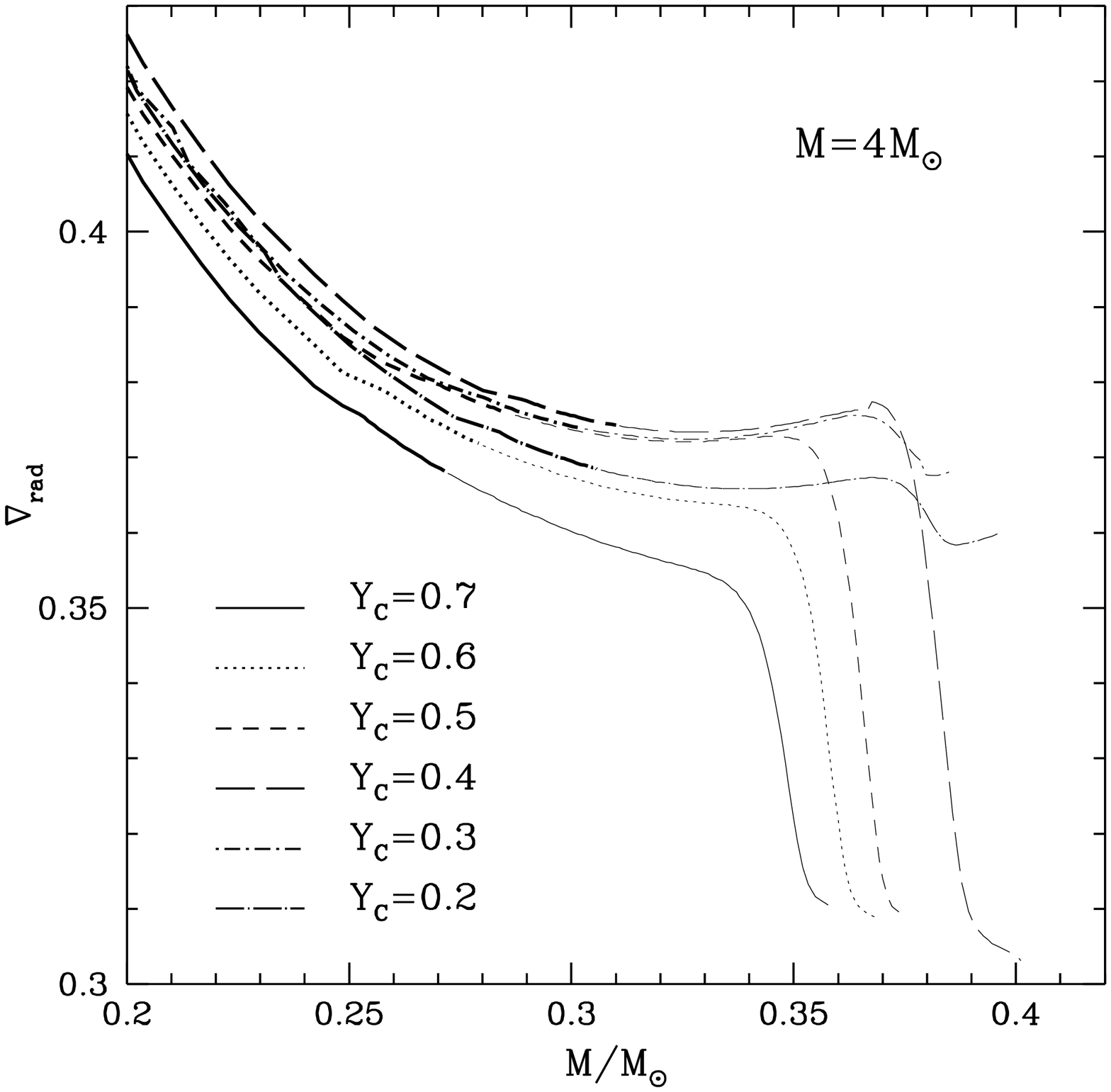}
           }
\caption{The internal distribution of opacity (left
         panel) and radiative temperature gradient 
         (right) within the core of a $4M_{\odot}$ model 
         at the same evolutionary stages of 
         Fig.~\ref{4morot}. We see that the increase of 
         the opacity coefficient at the border of the
         core leads eventually to the formation of a
         local minimum of $\nabla_{rad}$.}
   \label{4mopacity}
\end{figure*}

The two panels of Fig.~\ref{4mopacity} show the
internal variation of opacity (left panel) and
radiative gradient (right) for the same
6 models shown in Fig.~\ref{4morot}.
In the region of the $\log(\rho) - \log(T)$
plane covered by the core of the $4M_{\odot}$
model, i.e. $3.4 < \log(\rho) < 4$ and
$7.9 < \log(T) < 8.2$ (see the right panel of
Fig.~\ref{4morot}) we know that the dominant
contribution to the opacity is provided by
Thompson scattering, but with a non negligible
contribution by free-free transitions. This
latter term shows a ``thermodynamic'' dependence
of the form $\rho/T^{3.5}$ (which explains the
increase of the opacity when moving
from the centre of the star outwards) and a chemical
factor which increases for larger $^{12}C$
and $^{16}O$ abundances (Castellani et al. 1985). 
The two above terms, for a given mass 
shell, tend to balance each other,
because as He-burning proceeds the 
temperature increases and the helium 
mass fraction decreases,
so that the two contributions 
show an opposite behaviour: this is the reason
why for each value of the internal mass the
value of $\kappa$ inside the core is approximately
constant. The situation for the border of the 
mixed region, where we have a drop of the 
helium abundance and a maximum of the opacity,
is different, because its temperature is almost
constant as the evolution proceeds: this maximum
value of the opacity is therefore 
growing (see the left panel 
of Fig.~\ref{4mopacity}). It is this growing
contribution of the opacity which eventually
leads to the formation of a local maximum of the
radiative gradient, as can be seen in the right
panel of Fig.~\ref{4mopacity}.

\begin{figure*}
\centering{
\includegraphics[width=8cm]{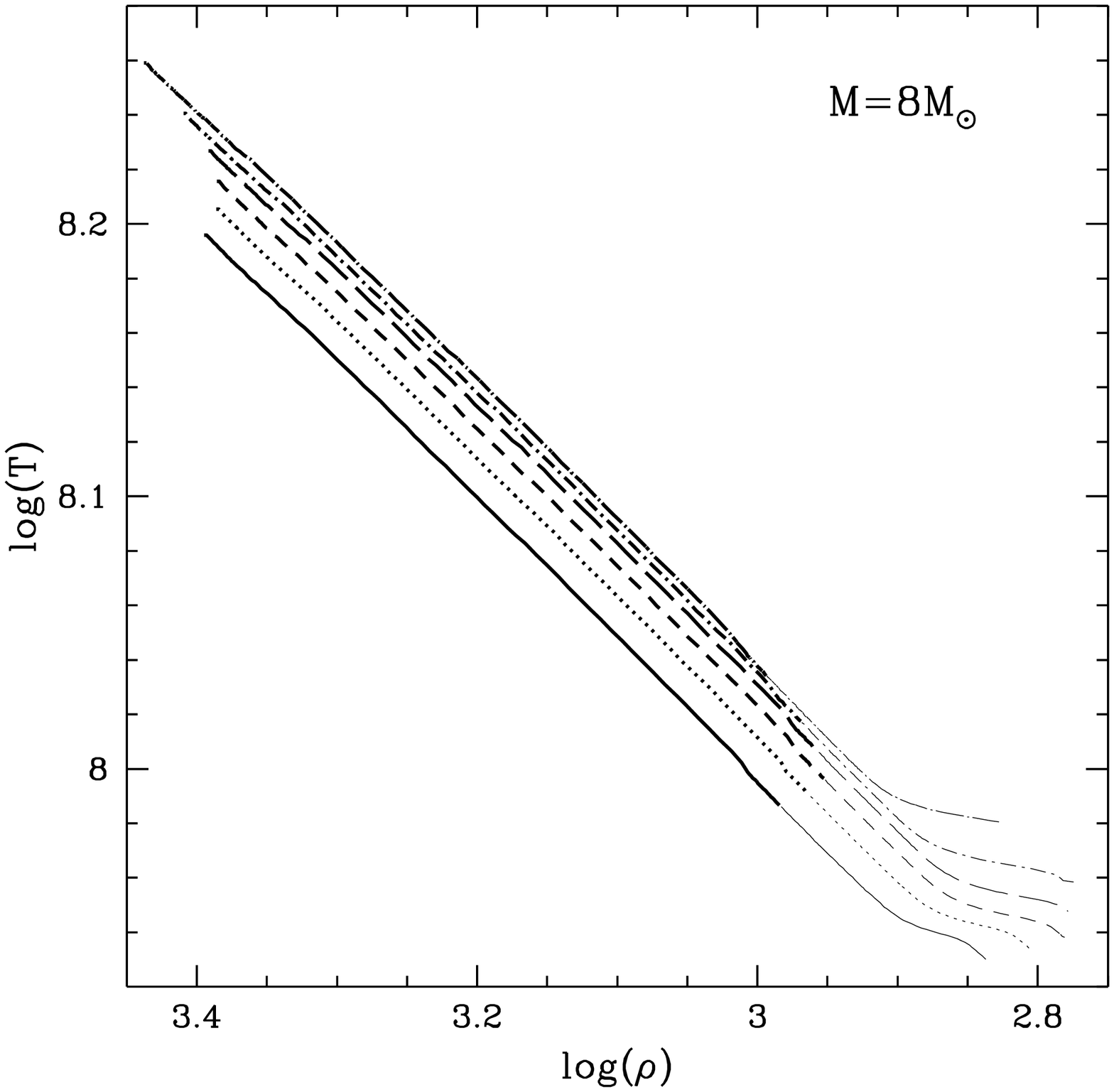}
\includegraphics[width=8cm]{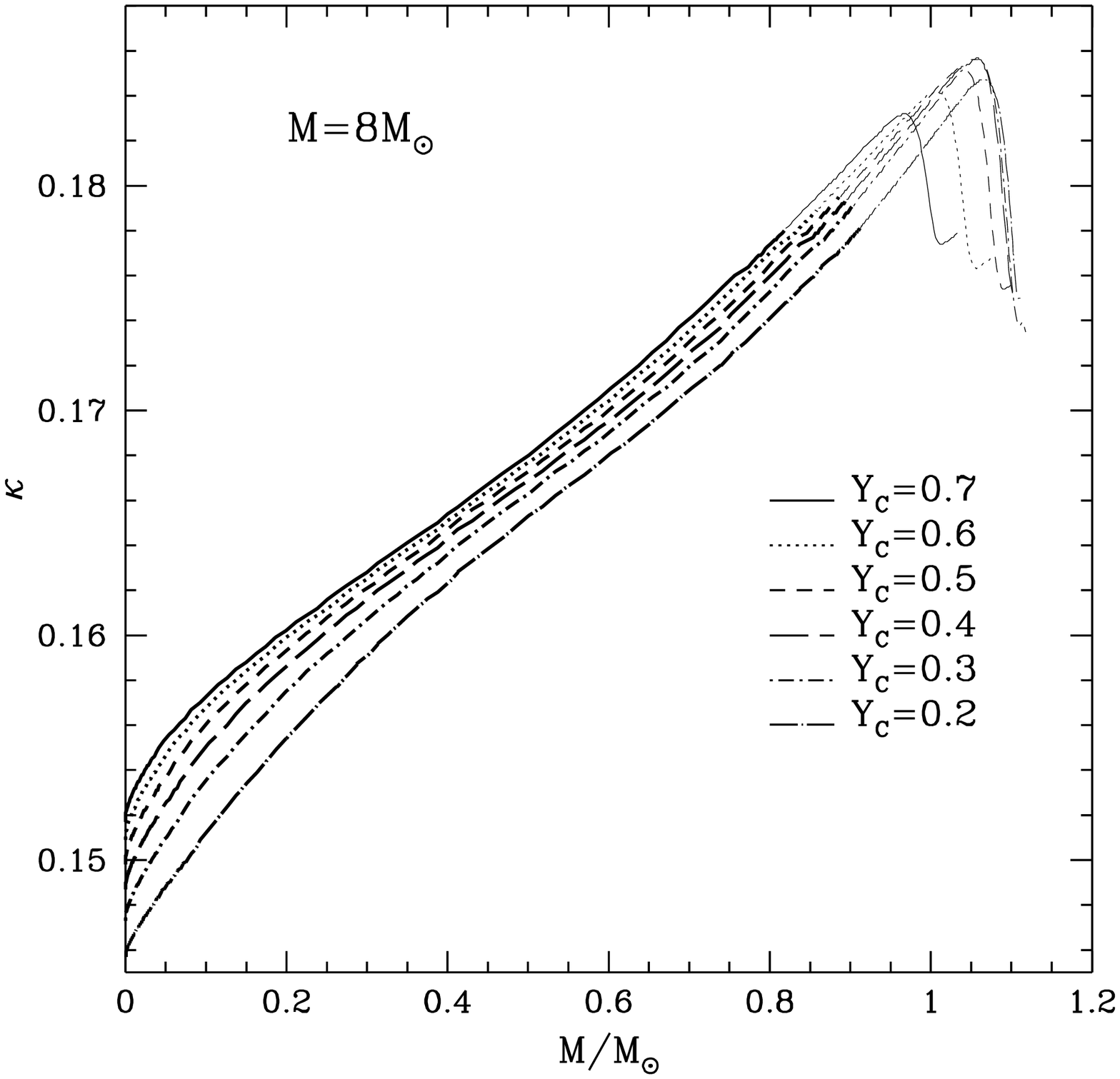}
           }
\caption{The structure of the core of a $8M_{\odot}$
         model during the He-burning phase at various
         evolutionary stages corresponding to some
         fixed values of the central mass fraction of
         helium. {\bf Right}: The $\log(\rho) - \log(T)$
         plane. {\bf Left}: The opacity coefficient.}
   \label{8mopacity}
\end{figure*}

From the above discussion it is clear that a
fundamental condition for the appearance of the
local maximum of the opacity is that the stellar
layers must be helium poor: this makes this
question relevant only in the final part of
this evolutionary phase, when the helium content
in the central layers drops to $Y_C < 0.2$.
This is confirmed by the smoothness of the solid
track of Fig.~\ref{4morot} in the early phases
of He-burning. 

Numerically, in a diffusive framework, the treatment of
the convective zone close to the central core is not
straightforward, because any exponential decay from 
a convective region so thin in mass is highly unreliable: 
we might decide in this case to stick to the Schwarzschild
criterion, and to ignore any overshooting, but this choice
is confirmed to trigger the afore mentioned discontinuities
when this small instability region vanishes.

To eliminate these irregularities we decided, in the case
when the convective shell is very close to the central core,
to ignore its presence for what concerns the calculation of
the velocity profile, i.e. to assume the same velocity field 
(and particularly the same exponential decay from the core) 
which we would have if all the stellar layers external to 
the core were radiatively stable. 

The result is the dotted track in the left panel
of Fig.~\ref{4morot}, which is much more smooth 
than the corresponding solid line. We also
verified that the global time of helium burning, as
well as the relative duration of the blue and
red phases of the clump, are practically identical
in the two cases, because any sudden increase of the
helium abundance in the central regions has the
effect of increasing the reaction rates of the
helium burning reactions, and this compensates
the larger mass fraction of helium, which would 
lead to a larger He-burning time. 

To understand why these sudden increases of the central 
helium mass fraction are not found in the evolution of 
the more massive models, we show in Fig.~\ref{8mopacity} the
internal structure of the core of the $8M_{\odot}$ model at 
various evolutionary stages identified by some fixed values of the 
central mass fraction of helium, in the $\log(\rho)-\log(T)$ 
plane (left panel) and in terms of the opacity profile (right panel).
If we compare the right panel of Fig.~\ref{4morot} 
with the left panel of Fig.~\ref{8mopacity} we 
see that the core of the $8M_{\odot}$ model
occupies a region in the $\log(\rho)-\log(T)$ plane
which is approximately at the same 
temperature as the $4M_{\odot}$ model, but at
lower densities: in this region the deviation
of the opacity coefficient from the Thompson
value are expected to be lower, and this
is confirmed by the left panel of 
Fig.~\ref{8mopacity}, where we can see that the
the values of $\kappa$ within the core are within
$0.15 < \kappa < 0.19$, to be compared to the much
wider range observed in the left-panel of
Fig.~\ref{4mopacity} (in making the comparison
between the two figures, note that in the
$4M_{\odot}$ case only the outer part of the 
core is reported, while in Fig.~\ref{8mopacity}
the whole central region is shown).

The maximum values of the opacity coefficient
found at the border of the convective core
($M \approx 1M_{\odot}$) of the $8M_{\odot}$ model
are therefore too low to cause the formation
of a local maximum in the $\nabla_{rad}$
profile, so that no convective zones in the
proximity of the core are found.

\subsection{The velocity field}
\begin{figure*}
\centering{
\includegraphics[width=8cm]{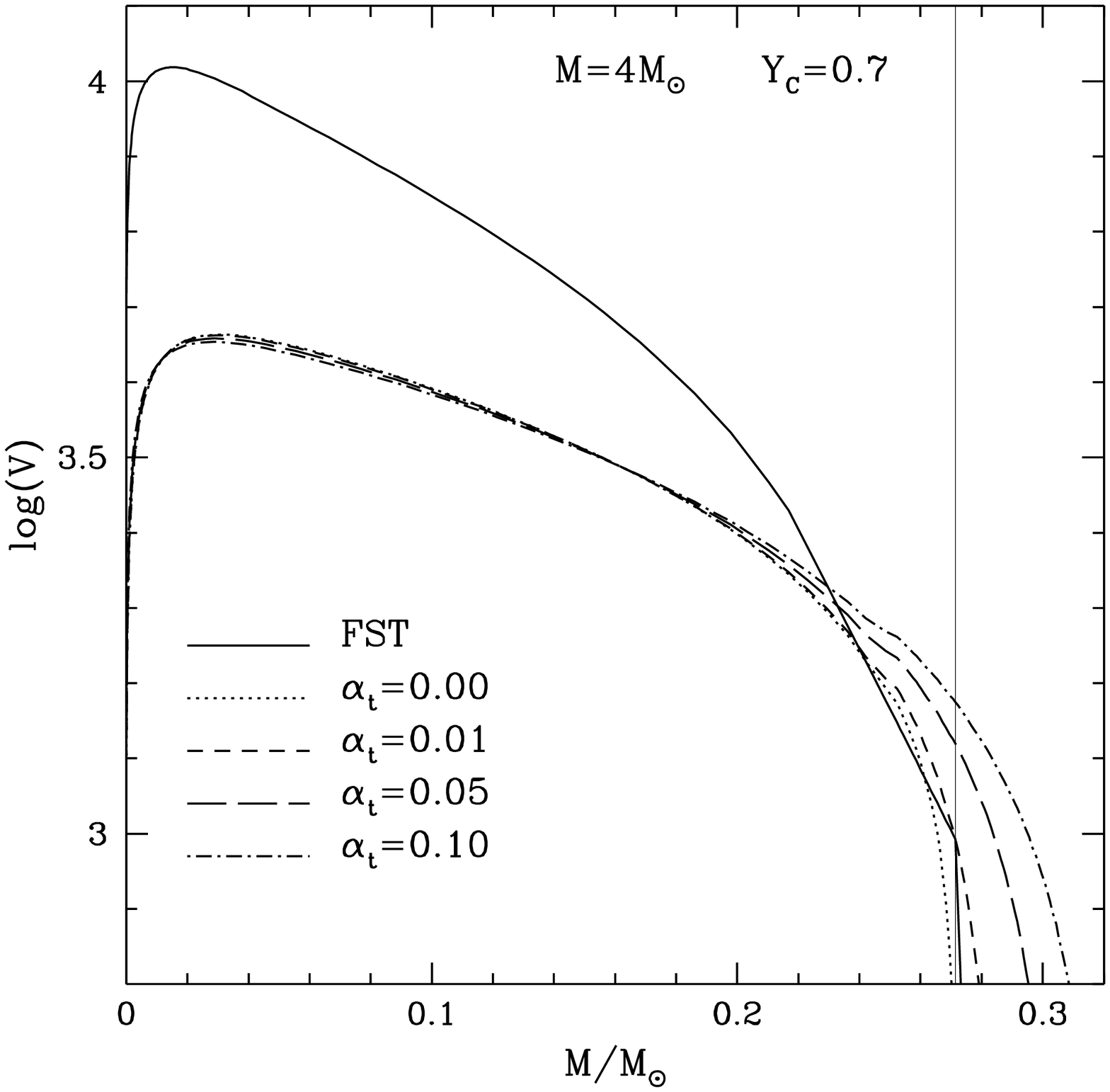}
\includegraphics[width=8cm]{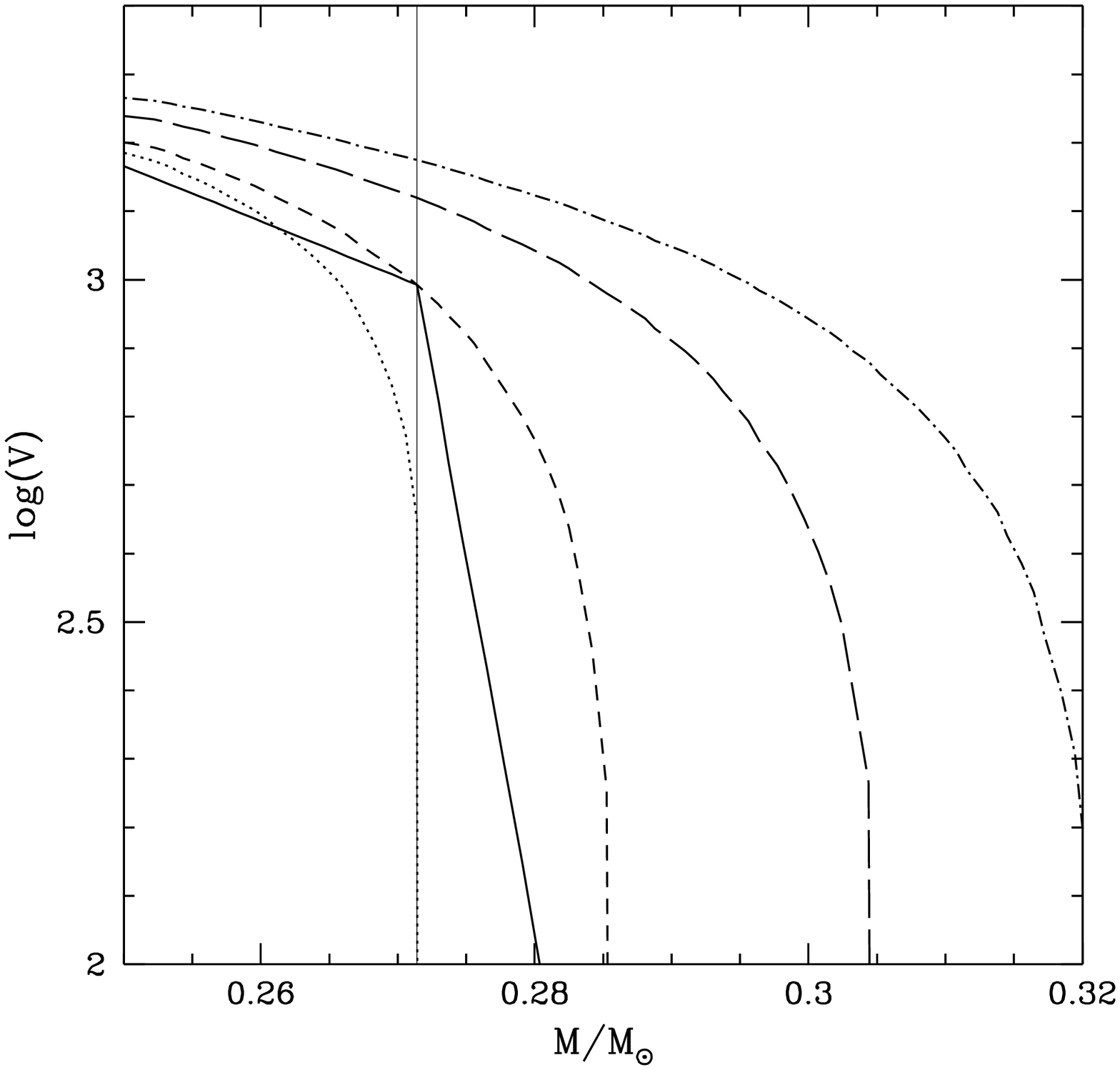}
           }
\caption
{Convective velocities inside the 
core of a $4M_{\odot}$ model during the phase
of core He-burning, when the central helium
mass fraction is $Y_C=0.7$. In the right panel
we report the whole convective core, in the
left we focus in the regions close to the
formal border, indicated by the thin vertical 
line. The solid line corresponds to the FST, the
remaining to the Kuhfu{\ss} case for four different
overshoot parameters in the range 
$\alpha_{t} = 0$--$0.10$.
}
  \label{veloc}
\end{figure*}

As we saw at the beginning of this section,
the rapidity with which helium is consumed 
within the core plays a fundamental role in
determining the time which IMS spend in the blue
region of the clump in the HR diagram. We will 
see in the following section that the time-scale
of the core He-burning phase is the main
difference between the diffusive models
and those calculated with the instantaneous
mixing approximation: these latter models,
burning helium more rapidly, stay very shortly
in the blue.

Within the diffusive framework, nuclear burning
and mixing are coupled via Eq.1. The diffusive
coefficient $D$ is found via Eq.2, and is 
dependent on the mean velocity of the convective 
eddies, $v$. A reliable estimate of $v$ is therefore
mandatory in determining correctly the importance 
of the diffusive term relatively to the nuclear
rates. Wrong values of the velocities, particularly
in the regions next to the border of the core, 
might lead to an incorrect extension of the 
region beyond the Schwarzschild border which is
fully homogenised, and, even more important for
our purposes, of the zone external to the core
where we still have some mixing, though on times
comparable to those typical of nuclear burning
in the interiors. 

In the FST context of turbulent convection, the
velocity is found by means of Eq.88 in Canuto
et al. (1996): we recall that in the FST model
the whole spectrum of eddies is taken into 
account, starting from the largest scales, of
dimensions comparable to the length of the
whole convective zone, to those typical of
the molecular dissipation processes. 

The FST model is local, in the sense that the 
convective gradient is calculated exclusively 
on the basis of physical quantities taken at the 
same locus where we calculate the gradient itself: this, 
unfortunately, neglects one of the essential 
features of the Navier-Stokes equations, i.e.
non locality. We therefore may wonder how the
velocities found in the FST context of convection
depend on the assumptions which
are at the basis of the model itself.

We therefore decided to compare our velocities
with those found by adopting a non-local theory,
developed by Kuhfu{\ss} (1986). 
Kuhfu{\ss} derives an equation for the specific
turbulent kinetic energy by proper spherically
averaging the first order perturbed Navier-Stokes
equations. The approach is similar to the models
by, e.g., Xiong (1981), Stellingwerf (1982,1984) and
Canuto (1997), and it shares with them the principal
difficulty to model the unknown correlation
functions of the fluctuating quantities. In the
Kuhfu{\ss} model this is done within the framework
of anelastic and diffusion type approximations
thereby introducing free parameters for every term
arising in the equation. Altogether we are left with
five free parameters. Kuhfu{\ss} fixes two of them
by matching the convective velocity and the convective
flux with the corresponding MLT values and we use
the same values here. The remaining parameters
consist of a mixing length parameter, which we set
close to the solar value of $1.6$ and an overshooting
parameter. The latter regulates the terms governing
the non-locality and it is the most relevant
parameter for the velocities at the Schwarzschild
boundary and the subsequent penetration of convective
eddies beyond this formal boundary.

Although we do not claim to have employed a final
theory of convection, the one from Kuhfu{\ss} is an
improvement over MLT since it allows predictions
of time-dependency and non-locality based on the
hydrodynamic equations. It is worth noting that
in the stationary, strictly local limit a cubic
equation similar to MLT is retained when the
Ledoux criterion is employed. In addition, the
Kuhfu{\ss} theory gives the same qualitative
behaviour for the temperature stratification in
the overshooting region as derived by Zahn (1991).
In the following, stationary solutions of the
Kuhfu{\ss} equation are calculated and compared
with the FST treatment.

We focus our attention on the same evolutionary
stages evidenced in Fig.~\ref{4morot},
and perform a detailed comparison between the
internal profile of the velocity provided by the
FST treatment with the values predicted by the
Kuhfu{\ss} theory for the same thermodynamic 
stratification. We stress here that our scope 
is neither to give a definite answer 
on how the convective 
eddies move in the convective core, nor which 
is their behaviour at the formal border; this 
is far beyond the scope of the present paper.
Our main goal is simply to test the robustness of the
velocities we use inside the inner parts of the
convective core up to the formal boundary,
comparing the results obtained by two completely
different approaches. Before going into the
details of such a comparison, we must say that
we are essentially interested in
stellar regions close to the formal border 
of the core fixed by the Schwarzschild criterion 
for the following reasons:

\begin{enumerate}
\item{In the deep interiors of the core, as we will
also show in the following section, the rapidity of
the convective motions is so high compared to the 
nuclear processes that it is possible to assume
that on a chemical point of view these regions
are completely homogenised.}

\item{In the proximity of the border the time
scale of convection increases, and the transport
of chemical becomes slower.}

\item{In the framework of extended convection with
exponential decay for the convective velocities,
as expressed in Eq.3, a key-role is played by
the velocity $v_b$ at the border of the core itself:
here, it is this quantity which eventually
determines the extension of the region beyond
the formal border where we find some chemical mixing.}
\end{enumerate}

\noindent
Fig.~\ref{veloc} shows the velocity profiles
inside the convective core of a $4M_{\odot}$
model, when the central mass fraction of helium
is $Y_C=0.7$. We focus on this mass, because
we will see that it is the model where the
differences between the diffusive and the
instantaneous schemes are more pronounced.

In the left panel we report the whole
convective core, while in the right panel we
show only the regions close to the border.
The solid line indicates the velocities found
through the FST model, while the other lines
refer to velocities calculated by the Kuhfu{\ss}
theory, each corresponding to a different value
of the overshooting parameter
$\alpha_{t}$. As can be seen in the right panel,
velocities in the model with $\alpha_{t}=0$,
which corresponds to strictly local convection,
terminate exactly at the Schwarzschild boundary.
Any overshooting is regulated by switching on the
non-local terms, i.e. $\alpha_{t}>0$. The exact
value of $\alpha_{t}$ is highly uncertain,
therefore we have chosen to calculate a few models
with $\alpha_{t}=\{0.005,0.01,0.05,0.10,0.20\}$.
The right panel of Fig.~\ref{veloc} shows that the
best match of the velocity at the formal
boundary between Kuhfu{\ss} model and FST
corresponds to a model with
$\alpha_{t}=0.01$ which is equivalent to small
overshooting of $0.06 H_p^b$. This is slightly
smaller than would be expected on other grounds,
since the best estimate for the overshooting distance
derived from isochrone fitting to colour-magnitude
diagrams of open clusters, e.g., by
Prather \& Demarque (1974), Maeder \& Mermilliod (1981),
Pols et al. (1998) and Demarque, Sarajedini, \& Guo (1994)
yield a canonical value of $0.23 H_p^b$ which would
correspond to $\alpha_{t}\approx 0.15$. In any case,
the velocities at the boundary deviate
at most by a factor of $1.6$ which can be
regarded as a close match considering that the
velocities are derived by completely different
approaches.

In the deep interior we see a qualitatively 
similar behaviour, with a maximum value which
is reached $\approx 0.02M_{\odot}$ away 
from the centre: we note that there is a 
difference of a factor of $\approx 2$ between 
the peak values, the FST model giving 
$v_{max} \approx 110$ m/s, to be compared to 
$v_{max} \approx 55$ m/s provided by the Kuhfu{\ss} 
modelling. 

In the proximity of the border, the 
two models provide essentially the same
result, i.e. $v_b \approx 10$--$15$ m/s, and this 
holds practically for all the Kuhfu{\ss} models,
independently of the overshooting parameter. 
From the right panel of Fig.~\ref{veloc},
we see that the values of the velocities found via the 
FST scheme plus the assumed exponential decay are 
comparable to those provided by the Kuhfu{\ss} models, 
with the closest match found for small overshooting 
parameters.

We could verify that the similarity of the 
results concerning the profile of velocity close
to the formal border of the core during the 
He-burning phase still holds, qualitatively and
quantitatively, until $Y_C > 0.4$. At later 
evolutionary stages the problems discussed 
in this section make the results obtained 
extremely sensitive to the way with which the 
small convective zone which develops away from
the core is treated, and how it is related to the
core itself. In this case a straight comparison
between the local and the non-local model is not
straightforward, because the results provided by the 
latter turn out to be much more strongly dependent 
than the previous phases on the overshooting parameter:
for the same thermodynamic stratification, we may
have either two distinct separated convective zones,
or the formation of a unique central instability
region, which, given the high values of the velocity
($v_{min} \approx 6$ m/s), will be fully homogenised.

The results in this case are much more uncertain.
Yet, we must recall that we are mainly interested
in the relative duration of the blue phase of
the core He-burning, which is determined by the 
rapidity with which helium is consumed in the core
before the bluest point of the track is reached:
therefore, our main interest is to evaluate the
robustness of the estimates of the velocities
close to the formal border of the convective core
during earlier phases, when $Y_C \geq 0.5$. 

We therefore find that in the models with masses
$M \approx 4M_{\odot}$, which we will see to be those
for which the results obtained depend critically
on the adopted scheme for mixing, the values of
the velocity found via the FST model in the regions
close to the border of the core for the
evolutionary phases preliminary to the bluest
point of the track are within a factor of $1.5$
consistent with those found 
by using a completely different approach, in which 
convection is treated non locally: though we do not
claim to have solved the problem of understanding
how the convective eddies are slowed down by the
opposition of the buoyancy forces, we are confident 
that our estimate is not far from the true value, 
which gives robustness to our results.

\section{The comparison between diffusive 
and instantaneous models}

In most of the modern evolutionary codes,
with only a few exceptions (see e.g.
Stancliffe et al. 2004),
mixing of chemicals inside the regions unstable 
to convection is treated as it was instantaneous, i.e.
the time scale of convection is assumed to
be much shorter than the time scales of all
the reactions included in the network, so
that it is assumed that the whole convective
region is instantaneously homogenised: this 
assumption allows to deal simultaneously with
nuclear burning and mixing, so that the 
whole instability zone is treated as if it
was a single mesh-point, for which average
cross-sections and chemical abundances are
found and used.

On a purely theoretical ground the two processes
should be considered simultaneously, and the
above simplification could be used and is 
expected to give results in agreement with the 
self-consistent treatment only if all the relevant 
nuclear reactions everywhere inside the 
convective region are much slower than the process
of mixing; we stress here that in order to have
similar results the above condition must hold also
in any extra-mixing region, i.e. in any region
beyond the formal border fixed by the Schwarzschild 
criterion where it is assumed that some sort
of mixing happens, despite the opposition of 
buoyancy.  

We calculated two sets of models, which were evolved
from the pre-MS phase up to the beginning of the
AGB:
\begin{enumerate}
\item{The diffusive models presented and discussed
in the previous section, with a parameter for
the exponential decay of velocity $\zeta=0.03$.}

\item{A set of models in which the mixing of
the chemical elements within the convective core
and in the extra-mixing region is 
assumed to be instantaneous. In this case
we used an overshooting distance of $0.2H_p$:
this latter choice is motivated by the fact that
this overshooting distance leads to results 
very similar to those of the diffusive models
with $\zeta=0.03$ in terms of the extension and
the duration of the main sequence of the single 
stars (VC05).}
\end{enumerate}

\begin{figure}
\includegraphics[width=8cm]{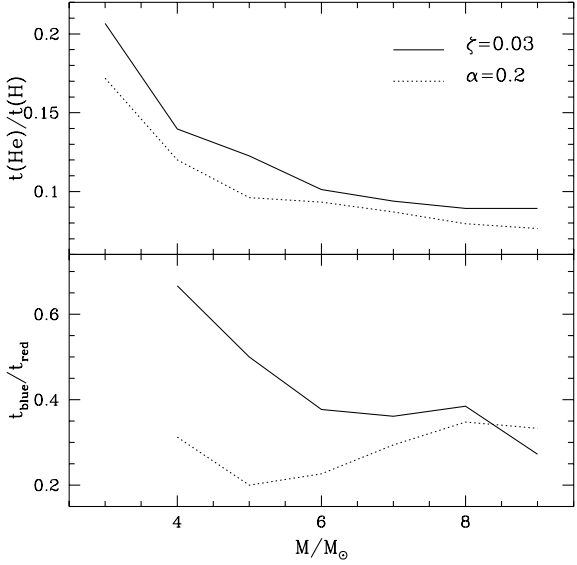}
\caption{{\bf Top}: The comparison of the ratio of
         the He-burning and the H-burning times for
         various masses between the diffusive 
         (solid line) and the instantaneous models 
         (dotted). {\bf Bottom}: The comparison of the
         ratio between the time spent in the blue and
         in the red region of the clump during the
         He-burning phase for stellar models of different
         mass calculate with the diffusive (solid) and the
         instantaneous (dotted) scheme for mixing.
         }
         \label{istvsdiff}%
\end{figure}

\noindent
Fig.~\ref{istvsdiff} shows for the two sets of
models and for each mass the ratio of the helium
burning time compared to the MS time (top panel)
and the ratio between the times spent, respectively,
in the blue and in the red part of the HR diagram
during the helium burning phase (bottom panel).

The top panel of Fig.~\ref{istvsdiff} shows that 
the He-burning time, relatively to the H-burning time,
is systematically lower in the instantaneous models.
From the bottom panel we can see
that the difference between the diffusive and the
instantaneous models in terms of $t_{blue}/t_{red}$
ratio decreases with mass, and eventually
vanishes for $M\approx 8M_{\odot}$. 

The different amount of helium still present in the
central core when the overall contraction begins
(i.e. when the surface convection is extinguished)
is the reason of the longer stay of the diffusive
models in the blue region of the clump, with respect
to the instantaneous models. In the $4M_{\odot}$ model
the central helium mass fractions when the excursion
of the track to the blue begins are $Y_C \approx 0.65$ 
and $Y_C \approx 0.40$, respectively, for the diffusive
and the instantaneous model. For the largest masses,
this difference decreases, and eventually vanishes
for $M=8M_{\odot}$, for which we find $Y_C \approx 0.30$
in both cases: this explains why the solid and the 
dotted line cross in the bottom panel of Fig.~\ref{istvsdiff}.

Before entering into the details of the models,
we also note from the same panel that 
$t_{blue}/t_{red}$ reaches a minimum value 
for $M\approx 7M_{\odot}$ for the diffusive models
and $M\approx 5M_{\odot}$ for the instantaneous set, and
is almost constant for higher masses. The occurrence
of the minimum marks approximately the border between 
the models for which the surface convection is 
extinguished when the core is still expanding 
($Y_C > 0.4$) and those for which the helium mass 
fraction at the beginning of the excursion towards
the blue part of the HR diagram is so low that the
central regions are beginning to contract to satisfy
the stellar energy demand: in the instantaneous case
the minimum is reached for a lower mass, because in
these models helium is consumed faster.
  
\subsection{The nuclear and mixing time-scales}
From the above discussion it is evident that the
reason for the diffusive scheme leading
to slightly longer He-burning times and to
generally longer fraction of time spent in the
blue region of the clump is readily explained by
a slower helium consumption in comparison with the
instantaneous case.
As can be seen in the bottom panel of 
Fig.~\ref{istvsdiff}, this difference is 
particularly relevant for the $4M_{\odot}$ model, 
so we temporarily focus our attention on this mass.
Qualitatively, we expect the same arguments to
hold even for slightly larger masses.

\begin{figure}
\includegraphics[width=8cm]{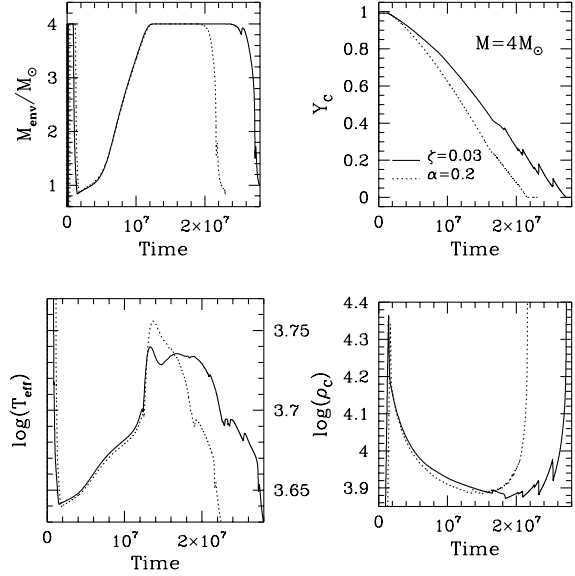}
\caption{The comparison between the evolutions of two
         models of $4M_{\odot}$ calculated with a diffusive
         (solid) and an instantaneous (dotted) scheme for
         chemical mixing. The four panels show the variation
         with the He-burning time of the base of the outer
         convective zone (left-upper panel), the central
         helium mass-fraction (right-upper panel), the
         effective temperature (left-bottom) and the central
         density (right-bottom).}
         \label{4confr1}%
\end{figure}

In the four panels of Fig.~\ref{4confr1} we compare the 
temporal evolution of some chemical and physical quantities
related to the evolution of the two models. The small
irregularities present in the diffusive model (solid
track) have been widely discussed in Sect.3.1. Here we 
only note that in the instantaneous model (dotted
track) such discontinuities in the temporal evolution
of the various physical and chemical quantities are
practically absent, because the small convective zone
which forms next to the formal border of central
convection (see Sect.3.1) is well inside the 
overshooting zone (whose width is $0.2H_p$), which 
is fully mixed in the instantaneous scheme.

In both cases we see that the surface convection 
is extinguished $\approx 12$ Myr after 
the beginning of helium burning (left-upper panel).
To this similarity between the times of the recession of
the convective envelope does not correspond an 
analogous behaviour of the helium burning: in the 
right-upper panel we see that helium consumption is faster in the
instantaneous model (dotted track), so that, when the
bluest point is reached, the central mass fraction of 
helium is $Y_C=0.65$ in the diffusive case, while it
is only $Y_C=0.4$ in the instantaneous model. As already
pointed out in the previous section, this lower
helium abundance determines an earlier start of 
the core-contraction, and a quick return of the
track to the red (see the bottom panels).

To resolve all doubts related to a possible
occurrence of any difference arising from
the physical and chemical
structure of the models at the beginning of
the He-burning phase, we calculated four evolutionary
sequences beginning from the same starting model,
at the tip of the giant branch, 
adopting a diffusive scheme for mixing with 
parameters $\zeta=0.02$ and $\zeta=0.03$,
and an instantaneous mixing approximation with
overshooting parameters $\alpha=0.2$ and $\alpha=0.3$.
The evolution from the pre-MS phase up to the
beginning of He-burning was calculated with
$\zeta=0.03$.

\begin{figure}
\includegraphics[width=8cm]{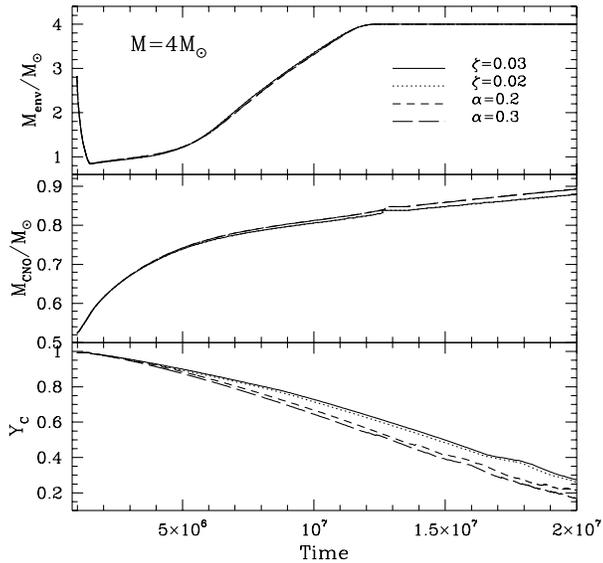}
\caption{The evolution from the tip of the giant branch
         of four $4M_{\odot}$ models calculated from the 
         same starter file. The solid and dotted lines
         refer to diffusive models differing in the $\zeta$
         coefficient, while the dashed and the dashed-dotted
         lines indicate the evolutions of two instantaneous
         models. The three panels report the variation of
         the mass of the base of the envelope (top), the
         mass where the peak of the CNO burning shell is
         located (middle) and the central mass fraction of
         helium.}
         \label{4confr2}%
\end{figure}

In the three panels of Fig.~\ref{4confr2} we compare 
the four evolutions.
We see that in all the sequences the rapidity
with which the surface convection recedes (top
panel) and the CNO burning shell moves outwards 
(middle panel) is the same, while in the instantaneous
models helium is burning faster: what is interesting
is that this qualitative difference between the two
sets of models holds independently of the values of 
the free parameters $\alpha$ and $\zeta$, and seems
to be more related to the details of the mixing scheme
adopted. All the models show the same evolution
for the outer layers of the star, starting
from the CNO burning shell, but the evolutions of the
core vary with the mixing scheme.

We stress that these differences are more relevant
here than in the context of H-burning, because in this
latter case we know that the core shrinks in mass
as H-burning proceeds, and this tends to level off
any difference.

A deeper inspection of the physical situation in
the interior of the core may help in understanding
the differences in the He-burning among the models.
As already shown in Fig.~\ref{4mopacity}, the 
$4M_{\odot}$ model develops a convective core 
(we refer only to the formal convective region,
excluding for the moment any extra-mixing zone)
whose extension in mass stabilises to 
$M_C \approx 0.29 M_{\odot}$ shortly after the 
beginning of He-burning. Within this region
the efficiency of convection is so high (due 
to the high densities) that the temperature gradient 
is practically adiabatic. The average velocity of 
the convective eddies reach a maximum value
of $v_{max} \approx 100$ m/s in the proximity of the 
centre of the star, and then declines to 
$v_b \approx 10$ m/s at points where
the formal border of the core is located
(see Fig.~\ref{veloc}). 
These velocities lead to a typical
time of mixing of the order of $\tau_{mix} \approx
2$ days in most parts of the core, which is
extremely short compared to the time scale of nuclear
burning by the $3\alpha$ and by the $^{12}C+\alpha$
reactions, which are of the order of 
$\tau_{nucl} \approx 10^5$ yr in the centre, and
$\tau_{nucl} \approx 10^{10}$ yr at the Schwarzschild
border (actually, the He-burning time gets progressively
shorter as He-burning proceeds due to the increase of 
the temperature, but the variations are at most by
a factor of ten, and do not change our main conclusions).
This discussion confirms that in the whole region
which is unstable to convective motions the use of the
instantaneous mixing approximation is consistent
with its basic assumption, i.e. convection is so
fast with respect to the nuclear processes that the
whole region is instantaneously homogenised.

The inclusion of overshooting has the effect of
increasing the size of the convective core: a
distance of $l_{ov}=0.2H_p$ makes the extension
in mass of the core to increase by $\delta M \approx 
0.04M_{\odot}$, and $l_{ov}=0.3H_p$ corresponds
to $\delta M \approx 0.08M_{\odot}$. In this extra-mixing
regions the velocities decrease to a few m/s, so that
the times scale of convection increase, up to
$\tau_{mix} \approx 10^5$ yr. Hence, in these regions
the rapidity with which convection tends to homogenise
the structure is comparable to the velocity with which
helium is burnt in the central layers, so that
the instantaneous mixing approximation does not
hold. In this case, if the diffusive approach is
used, further helium from the outer layers of the 
star is still added to the central regions, but
this takes a longer time when compared to the
instantaneous scheme, so that the helium burning
process takes longer. The extension of the extra-mixing
region where the instantaneous mixing approximation
still holds depends on the details of the velocity
decay from the border, hence on the value of
$\zeta$ which is used, but, with the exception of
extremely low values of $\zeta$, there will be in
any case a region which is assumed to be fully
homogenised in the instantaneous case, which is
mixed on longer time-scale in the diffusive
approach. This is the ultimate reason of the
different qualitative behaviour of the instantaneous
and the diffusive models which can be seen in 
Fig.~\ref{4confr2}.

\begin{figure}
\includegraphics[width=8cm]{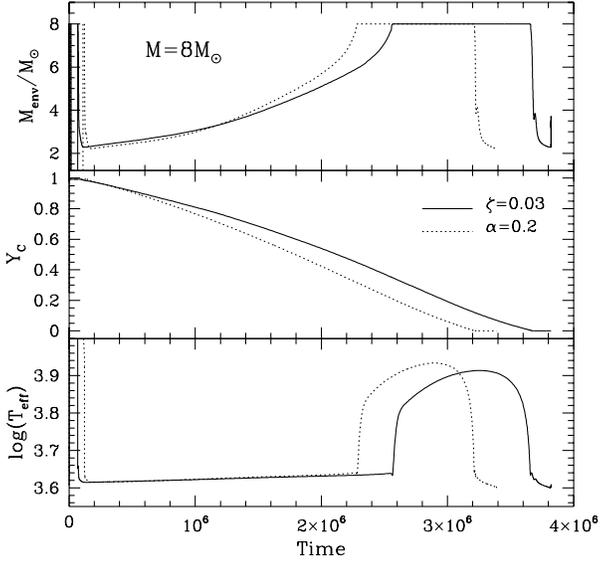}
\caption{The comparison between the evolutions of two
         $8M_{\odot}$ models calculated with a diffusive
         (solid line) and an instantaneous scheme (dotted)
         for chemical mixing inside the convective core. 
         The three panels show the variation with time of
         the mass at which the base of the external 
         envelope is located (top), the central helium
         mass fraction (middle) and the effective 
         temperature (bottom).
          }
         \label{8confr}%
\end{figure}

The same qualitative arguments still hold if we 
consider larger masses. In the $8M_{\odot}$ modes,
for example, the slightly different thermodynamic
conditions of the core favour convective velocities
which are larger by a factor of $\approx 2$ with respect
to the $4M_{\odot}$ case, but the larger extension
of the core makes the mixing time scales extremely
similar in the two cases. The larger temperatures
in the $8M_{\odot}$ case make the nuclear time
scale shorter, but the difference is contained
within one order of magnitude. Therefore, even in this
case we expect the presence of a region out of the
convective border which is only partly mixed in the
diffusive case, which makes the whole He-burning 
phase longer. This is confirmed by looking at 
Fig.~\ref{8confr},
where we can see the two $8M_{\odot}$ calculated 
with the diffusive (solid track) and the instantaneous
(dotted line) scheme for mixing differ in the He-burning
times, the instantaneous model being faster. At 
difference with the $4M_{\odot}$ model we note
from Fig.~\ref{8confr} that in this case also the times for
the recession of the convective envelope are different,
so that at the beginning of the excursion of the
track to the blue the two models have approximately
the same amount of helium left in the core: this
acts in favour of a similar fraction of time spent
in the blue part of the HR diagram during He-burning,
and is the reason why the gap between the ratio
of the time spent in the blue and red part of the
clump decrease with mass. 

We may conclude that, if we are interested in a 
correct estimate of the He-burning time of IMS, 
and particularly to the relative duration of 
the blue and the red phase, the use of the 
diffusive scheme for chemical mixing is recommended for models with mass 
$M \leq 6M_{\odot}$. The instantaneous mixing scheme would underestimate 
the total He-burning time, and the duration of the stay of the track 
in the blue region of the clump. Only for more massive models 
the instantaneous mixing scheme leads to results 
similar to those found by using the diffusive approach.

\subsection{The clump population of open clusters}
The discussion of the previous subsection shows that 
a particular range of stellar masses exists for which 
the relative duration of the red and blue phase of
He-burning show a dependency on the scheme used
to deal with mixing and burning in the central core.

To understand in which cases the description of mixing
in the central regions is relevant for the theoretical 
interpretation of observational photometric results
we constructed isochrones for the
instantaneous and the diffusive models corresponding to
ages in the range $50 - 250$ Myr both in the theoretical
and in the observational plane. We ruled out older ages
because in those cases the extension of the clump, in
terms of the excursion of the track to the blue, 
is very short, rendering extremely difficult to 
identify a red and a blue region (see the track
of the $3M_{\odot}$ model in Fig.~\ref{HRdiagram}). 
To construct the 
isochrones we followed the scheme suggested by Pols 
et al.(1998); the transformations from the theoretical 
to the observational plane, for that concerning the 
$B-V$ colours and the bolometric corrections, were 
performed by using the tables by Castelli et al.(1997).
For each isochrone we performed a numerical simulation on
the observational plane.

Such numerical simulations were built following the 
procedure already described in VC05, that we briefly recall in
the following. First, each selected isochrone
was populated with a random distribution of masses; then,
for each extracted mass, we obtained by linear interpolation
the corresponding values of LogL and LogTe, as well as
magnitudes and colours in the observational plane. Once
we populated a given isochrone, we were
able to derive the B/R value, by simply counting
the number of stars that fall within {\it appropriate boxes} (see below) in
the CM diagram.
 
To obtain a reliable estimate of B/R, 
we carefully selected different boxes for isochrone of
different ages, taking into account, in particular, the 
specific magnitude and colour extension of the ``blue loop''. 
We decided not to overlap``blue'' and ``red'' boxes, 
in order to make our statistics more robust against
small fluctuations in colour. To provide an estimate 
of the statistical fluctuation related to the random 
extraction of star masses, we performed ten simulations 
for each age, varying only the seed for random number 
generation, thereby deriving the mean B/R and the 
corresponding standard deviation for each set
of simulations (in this context, small values of 
standard deviations can be regarded as an indication 
of the goodness of our choice of the specific shape 
of blue and red boxes).

\begin{figure}
\includegraphics[width=8cm]{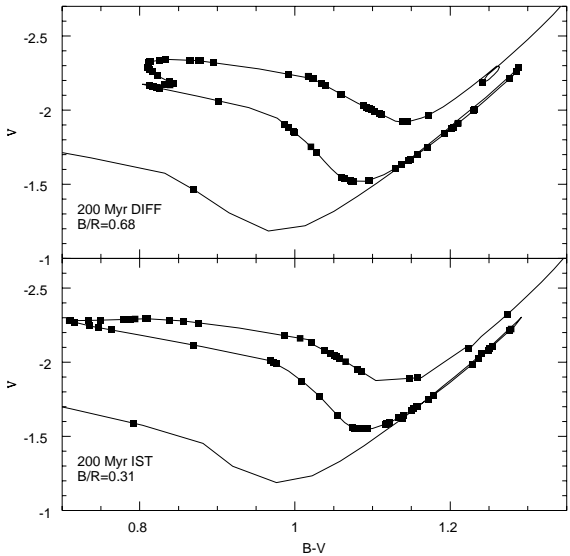}
\caption{The solid tracks in the two panels represents the theoretical 
         isochrones corresponding to an age of 200 Myr calculated
         with a diffusive (top) and an instantaneous (bottom) scheme
         for chemical mixing. The dots superimposed on the isochrones
         re the results of our numerical simulations.
          }
         \label{simul200}%
\end{figure}
\begin{figure}
\includegraphics[width=8cm]{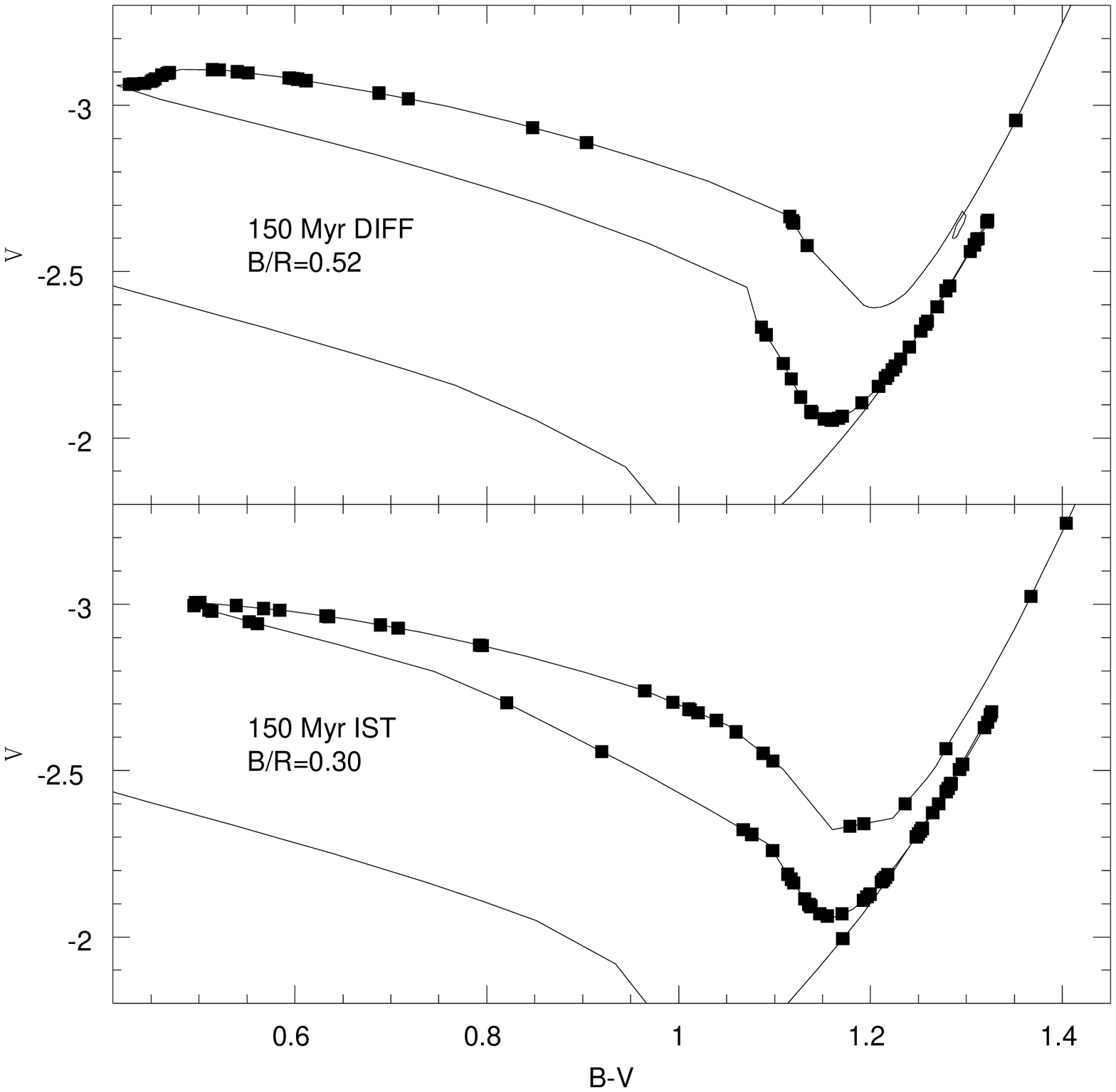}
\caption{The same as Fig.~\ref{simul200}, but for an
         age of 150 Myr.
          }
         \label{simul150}%
\end{figure}
\begin{figure}
\includegraphics[width=8cm]{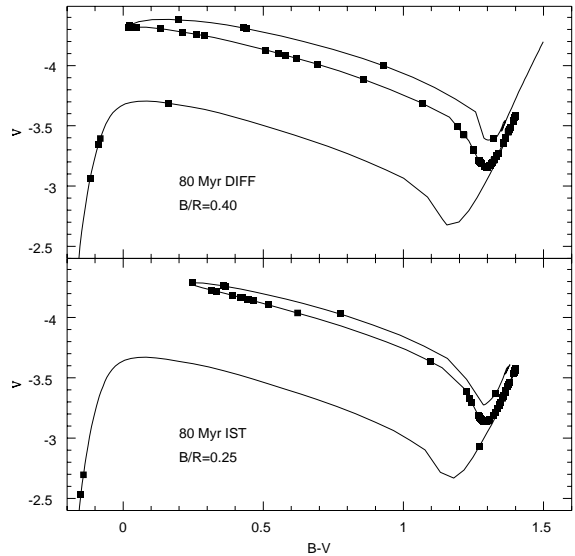}
\caption{The same as Fig.~\ref{simul200}, but for an
         age of 80 Myr.
          }
         \label{simul80}%
\end{figure}

We show three of such simulations in Figs.~\ref{simul200},
\ref{simul150} and \ref{simul80}. We focus our attention
on the region of the clump; for clarity 
reasons all the stars in the H-burning phase were 
omitted from the figures. For the statistical
analysis performed to be relevant, we supposed that the 
clump contains a considerable large number of stars 
($\approx 50$), as is the case, for example, for NGC1866 
(Testa et al. 1999). In all cases we see that 
it is possible to identify two distinct regions in the 
diagram where He-burning takes place: the excursion of 
the track to the blue, and the return to the red, 
are very quick, rendering extremely small the probability
of finding stars there.

As expected on the basis of the previous discussions, 
the differences between the 
predictions of the two sets of models are more relevant 
for the ages at which the typical mass of the He-burning 
stars is $M \approx 4M_{\odot}$ (see the bottom panel of 
Fig.~\ref{istvsdiff}), i.e. 200 Myr; we see from 
Fig.~\ref{simul200} that the $B/R$ ratio is 
$\approx 0.7$ for the diffusive case, and is only 
$\approx 0.3$ for the instantaneous models. These differences
diminish for younger ages (hence, larger clump masses): 
for $t=150$ Myr (see Fig.~\ref{simul150}), for which
the typical mass of the stars in the clump is 
$M \approx 4.5M_{\odot}$, the instantaneous $B/R$ is 
the same as before, i.e. $\approx 30\%$, while the 
diffusive value drops to $\approx 50\%$;

For younger ages the mass of the stars involved shifts 
into the range of values for which, independently of the 
details of the scheme followed to couple mixing and 
nuclear burning, the duration of the track in the blue 
is shortened by the scarcity of helium in the 
central regions: the simulation shown in 
Fig.~\ref{simul80}, which refers to an age of $80$ 
Myr (and to a mass on the clump of $\approx 6M_{\odot}$) 
demonstrates that the gap between the results of the 
two schemes reduces further, the predicted
$B/R$ ratios being $\approx 40\%$ and $\approx 25\%$,
respectively, for the diffusive and the instantaneous
models; our numerical simulations performed for younger
ages show that the differences between the predictions
concerning the $B/R$ ratio for the two classes of models
become too low to be statistically relevant.

We may therefore conclude this analysis by noting that, for
a metallicity typical of the LMC, the use of a diffusive 
approach, when compared to the instantaneous case,
leads to predict a larger population of He-burning 
stars in the blue part of the region of the clump, but 
these differences are limited to ages in the range 
$80 - 200$ Myr, for which the typical masses in the 
He-burning phase are $4M_{\odot} \leq M \leq 6M_{\odot}$; 
for older ages the extension of the clump itself is so 
short that it does not allow to identify two distinct 
regions, while for younger ages the differences between 
the predictions of the two sets of models are not 
statistically relevant.

\section{Conclusions}
We present detailed evolutionary models 
which focus on the 
core Hydrogen- and Helium-burning phases of intermediate
mass stars, with masses up to $M=9M_{\odot}$; the present
investigation employs a metallicity typical of the LMC,
i.e. $Z=0.008$.

We compare carefully the results obtained with the traditional
instantaneous mixing approximation 
when applied to chemical mixing
and nuclear burning in the convective core, and those found
by treating the two mentioned processes simultaneously, in a
self-consistent way, with a diffusive approach.

During the H-burning phase no relevant differences 
are found in the results between the two schemes for mixing,
because the convective core of these stars 
tends to shrink in mass; given a value of the 
mass of the star, the duration of this phase is almost 
independent of the mechanism chosen to deal with mixing.

In contrast to the H-burning phase, we find different 
results during the later phase of He-burning
when the diffusive scheme is used. Helium from
the outer layers, in the proximity 
of the formal core defined by the Schwarzschild
criterion is similarly efficient but 
slower mixed into the central regions in comparison
to the instantaneous models, and 
this leads to two important differences:

\begin{enumerate}
\item{For a given mass, the duration of the entire 
He-burning phase, and consequently the $t(He)/t(H)$ 
ratio, is longer in the diffusive models.
}
\item{In the diffusive models the fraction of the 
time spent in the blue region of the clump is
larger compared to the corresponding instantaneous 
models; this is the consequence of the more rapid  
helium consumption in the instantaneous case, which, 
as soon as the track reaches the bluest point in the HR
diagram, favours an early contraction of the core and 
a return of the track to the red.
}
\end{enumerate}
Both the above effects are particularly relevant for models
with mass $M \leq 6M_{\odot}$, because for higher masses the
larger mass of the envelope (which triggers a later exhaustion
of surface convection) and the more rapid burning of helium
forces the track to return quickly to the red.
Therefore, in the interpretation of
the stellar population of open clusters, 
the differences introduced by the use of the 
diffusive approach are relevant for stellar models in 
the range $3.5M_{\odot} \leq M \leq 6M_{\odot}$,
and for ages in the range 
$80$ Myr $\leq t \leq 200$ Myr.

These results prove to hold independently of all 
the uncertainties which characterise the choice 
of the diffusion coefficient entering into the 
diffusion equation, which ultimately depends on 
the velocity profile within and out of the central 
region: in the proximity of the core the values 
of the velocities found via the local FST model 
for turbulent convection turned out to be
extremely similar to those obtained via a 
non-local theory, and this holds independently 
of all the free parameters entering both treatments. 
The differences between the diffusive and the 
instantaneous models in the afore mentioned range 
of mass is essentially related to the way with 
which helium is mixed from the outer layers into 
the central regions, which is governed by an 
exponential decay of convective velocities.
The choice of parameters (e.g. the assumed 
distance of the exponential decay or the extension 
of the overshooting region in the instantaneous 
models) associated with a selected mixing scheme 
turns out to be largely insignificant.

\end{document}